\documentclass[acmtog,screen]{acmart}

\usepackage{booktabs}

\usepackage{algorithm}
\usepackage{algpseudocode}
\usepackage{multirow}

\usepackage{xcolor}

\usepackage{bbold}

\usepackage{pgfplots}
\usepackage{pgfplotstable}
\usepgfplotslibrary{groupplots}
\pgfplotsset{
    width=0.3\linewidth,
    compat=1.9,
}

\definecolor{my-indigo}{HTML}{5856D6}
\definecolor{my-tianyi}{HTML}{66CCFF}
\newcommand{\AlgSymbol}[1]{\textcolor{my-indigo}{{#1}}}
\newcommand{\AlgComment}[1]{\textcolor{my-tianyi}{{#1}}}

\definecolor{my-color-near}{HTML}{5ACDFA}
\definecolor{my-color-middle}{HTML}{148EFF}
\definecolor{my-color-far}{HTML}{6361F2}
\newcommand{\COLORNEAR}[1]{\textcolor{my-color-near}{{#1}}}
\newcommand{\COLORMIDDLE}[1]{\textcolor{my-color-middle}{{#1}}}
\newcommand{\COLORFAR}[1]{\textcolor{my-color-far}{{#1}}}
\definecolor{my-color-ours}{HTML}{009417}
\newcommand{\COLOROURS}[1]{\textcolor{my-color-ours}{{#1}}}

\definecolor{my-color-footprint-orange}{HTML}{FFA914}

\copyrightyear{2025}
\acmYear{2025}
\setcopyright{acmlicensed}
\acmConference[SIGGRAPH Conference Papers '25]{Special Interest Group on Computer Graphics and Interactive Techniques Conference Conference Papers}{August 10--14, 2025}{Vancouver, BC, Canada}
\acmBooktitle{Special Interest Group on Computer Graphics and Interactive Techniques Conference Conference Papers (SIGGRAPH Conference Papers '25), August 10--14, 2025, Vancouver, BC, Canada}
\acmDOI{10.1145/3721238.3730602}
\acmISBN{979-8-4007-1540-2/2025/08}

\setcitestyle{square}

\begin{document}

\title{Virtualized 3D Gaussians: Flexible Cluster-based Level-of-Detail System for Real-Time Rendering of Composed Scenes}

\author{Xijie Yang}
\orcid{0009-0009-3076-2595}
\affiliation{
  \institution{Zhejiang University}
  \city{Hangzhou}
  \country{China}
}
\affiliation{
  \institution{Shanghai Artificial Intelligence Laboratory}
  \city{Shanghai}
  \country{China}
}
\email{yangxijie@zju.edu.cn}

\author{Linning Xu}
\orcid{0000-0003-1026-2410}
\affiliation{
    \institution{The Chinese University of Hong Kong}
    \city{Hong Kong}
    \country{China}
}
\email{linningxu@link.cuhk.edu.hk}

\author{Lihan Jiang}
\orcid{0009-0001-2899-273X}
\affiliation{
  \institution{University of Science and Technology of China}
  \city{Hefei}
  \country{China}
}
\affiliation{
  \institution{Shanghai Artificial Intelligence Laboratory}
  \city{Shanghai}
  \country{China}
}
\email{jianglihan@mail.ustc.edu.cn}

\author{Dahua Lin}
\orcid{0000-0002-8865-7896}
\affiliation{
    \institution{The Chinese University of Hong Kong}
    \city{Hong Kong}
    \country{China}
}
\affiliation{
  \institution{Shanghai Artificial Intelligence Laboratory}
  \city{Shanghai}
  \country{China}
}
\email{dhlin@ie.cuhk.edu.hk}

\author{Bo Dai}
\orcid{0000-0003-0777-9232}
\affiliation{
  \institution{The University of Hong Kong}
  \city{Hong Kong}
  \country{China}
}
\affiliation{
  \institution{Feeling AI}
  \city{Shanghai}
  \country{China}
}
\email{bdai@hku.hk}

\renewcommand{\shortauthors}{Xijie Yang et al.}


\begin{abstract}
3D Gaussian Splatting (3DGS) enables the reconstruction of intricate digital 3D assets from multi-view images by leveraging a set of 3D Gaussian primitives for rendering. Its explicit and discrete representation facilitates the seamless composition of complex digital worlds, offering significant advantages over previous neural implicit methods. 
However, when applied to large-scale compositions, such as crowd-level scenes, it can encompass numerous 3D Gaussians, posing substantial challenges for real-time rendering.
To address this, inspired by Unreal Engine 5's Nanite system, we propose Virtualized 3D Gaussians (V3DG), a cluster-based LOD solution that constructs hierarchical 3D Gaussian clusters and dynamically selects only the necessary ones to accelerate rendering speed.
Our approach consists of two stages: (1) Offline Build, where hierarchical clusters are generated using a local splatting method to minimize visual differences across granularities, and (2) Online Selection, where footprint evaluation determines perceptible clusters for efficient rasterization during rendering.
We curate a dataset of synthetic and real-world scenes, including objects, trees, people, and buildings, each requiring 0.1 billion 3D Gaussians to capture fine details.
Experiments show that our solution balances rendering efficiency and visual quality across user-defined tolerances, facilitating downstream interactive applications that compose extensive 3DGS assets for consistent rendering performance.
\end{abstract}




\begin{teaserfigure}
    \centering
    \includegraphics[width=\linewidth]{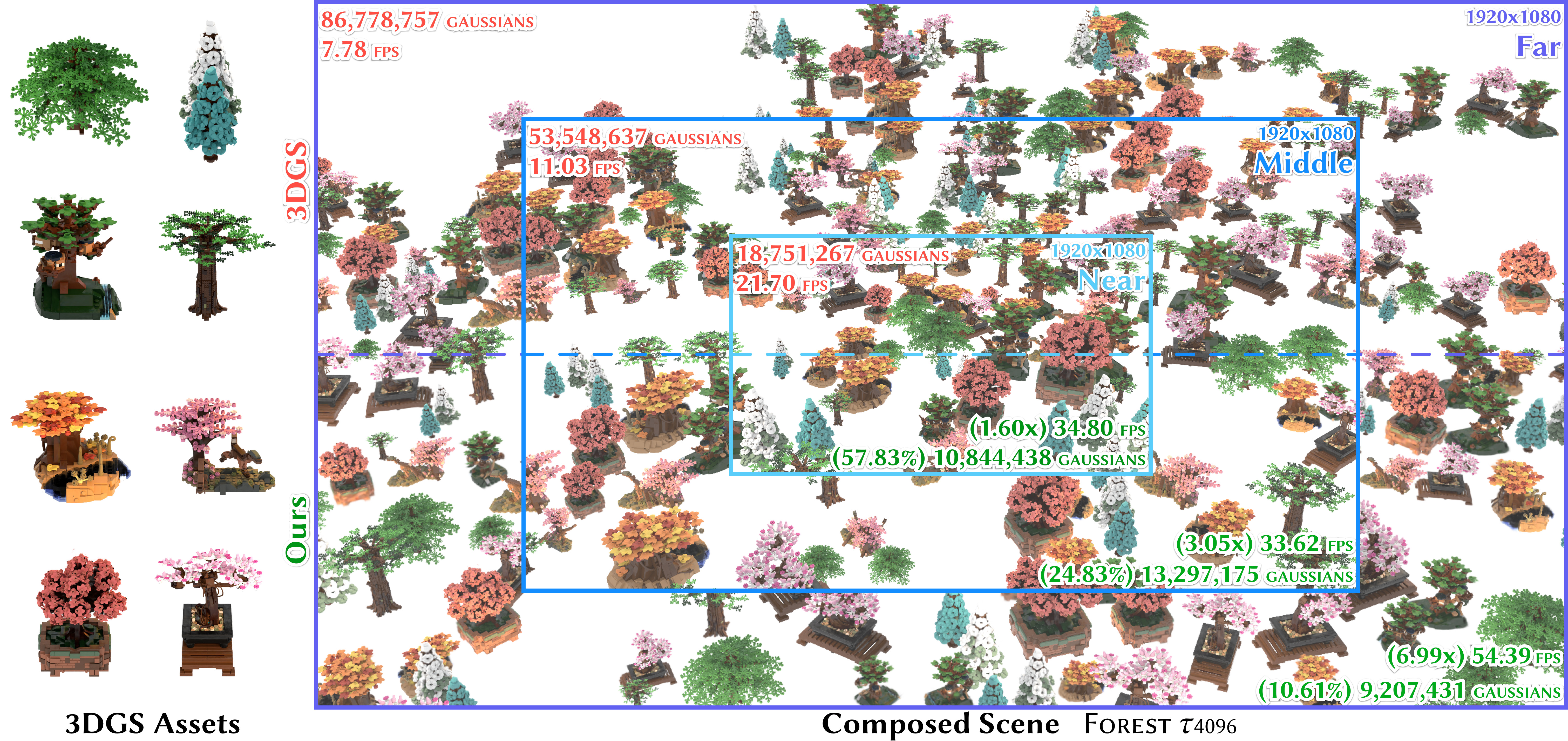}
    \caption{\label{fig:teaser}
        \textit{Left}: Collection of digital \textbf{3DGS assets}.
        \textit{Right}: Optimized 3D Gaussians (\COLOROURS{Ours}) in \textbf{composed scene} \textsc{Forest} with \textbf{adaptive LOD selection} based on distance, resolution, and user-defined tolerance, minimizing overhead while maintaining visual quality. We report number (percentage) and FPS (acceleration rate) of selected 3D Gaussians at various distances (\COLORNEAR{Near}, \COLORMIDDLE{Middle}, and \COLORFAR{Far}), showing consistent \textbf{real-time performance} compared to vanilla 3DGS composition.
    }
    \Description{teaser}
\end{teaserfigure}


\maketitle


\section{INTRODUCTION\label{sec:introduction}}

3D Gaussian Splatting (3DGS)~\cite{gs} reconstructs objects and scenes with explicit 3D Gaussian primitives from multi-view images, achieving high visual fidelity and real-time rendering, offering multiple benefits compared to previous NeRF-based methods~\cite{nerf}.
The reconstructed objects and scenes in 3D Gaussian representation, which we refer to as \textit{3DGS assets}, can be further used by digital artists to compose \textit{digital worlds}.

While individual 3DGS assets consisting of millions of explicit 3D Gaussians can be rendered in real-time with the GPU rasterizer~\cite{gs}, rendering a complex, large-scale digital world with numerous assets and 3D Gaussian in real time remains a significant challenge for interactive applications.
This limitation arises because 3DGS-based methods rely solely on visibility-based filtering to select the visible primitives, assuming that these primitives share the same level of importance.
As a composed scene grows with more 3DGS assets, primitives in the camera frustum become redundant and overlapping, causing a low rendering speed.
To solve this problem, a Level-of-Detail (LOD) system is necessary to control the rendering overhead dynamically under given rendering resolution and camera distance.

Inspired by the Nanite Virtualized Geometry system~\cite{nanite} in Unreal Engine 5, 
which efficiently handles a large amount of geometric detail through dynamic LOD streaming,
we propose Virtualized 3D Gaussians (V3DG), an LOD system for real-time rendering of a large number of 3D Gaussians in composed scenes. Specifically, our system comprises two stages: an \textit{offline build stage} and an \textit{online selection stage}.
(1) The offline build stage acts as a one-time post-processing step for each asset. Each 3DGS asset is first split into clusters, forming the finest layer. Then, these clusters are simplified using a novel \textit{local splatting} strategy to build coarser layers with minimal difference in a \textit{tree} structure, establishing parent-child relationships across levels of detail for selection.
The multi-level cluster data is stored as a \textit{bundle} for the selection stage.
(2) The online selection stage is embedded in each rendering process to lower rendering overhead.
Before the rasterization of 3D Gaussians, 
clusters from all bundles are projected onto the 2D screen space to derive their \textit{footprints} based on the camera view, measuring the level of detail for each cluster.
Clusters at appropriate levels of detail are then selected to meet a predefined \textit{footprint tolerance}, ensuring only necessary 3D Gaussians are rasterized further, which matches the detail of 3D Gaussians with the given rendering conditions, achieving faster rasterization and anti-aliasing effects.

We further curate a dataset consisting of four composed scenes with thousands of detailed 3DGS assets, covering both synthetic and real-world data, to better evaluate our method and advance the field.
Comprehensive experiments show that 3DGS struggles with real-time rendering of these challenging scenes, while our approach boosts rendering speeds under predefined tolerances, preserving visual quality and incorporating anti-aliasing for enhanced comfort. Users can also balance quality and speed by adjusting tolerances.

To summarize, our contributions are as follows:
\begin{itemize}
\item A post-processing cluster-based LOD system for rendering approximately 100 million 3D Gaussians in digital worlds in real-time, with a build stage constructing hierarchical clusters and a selection stage that filters necessary clusters.
\item A local splatting method that simplifies clusters across layers with minimal visual difference agnostic to training views.
\item A selection strategy based on footprint evaluation, achieving high-quality rendering with the minimally required number of 3D Gaussians.
\item A large-scale dataset with composed scenes containing various synthetic and real-world assets for evaluation of LOD system on 3D Gaussians.
\end{itemize}

\section{RELATED WORK}

\begin{figure*}[t]
    \centering
    \includegraphics[width=\linewidth]{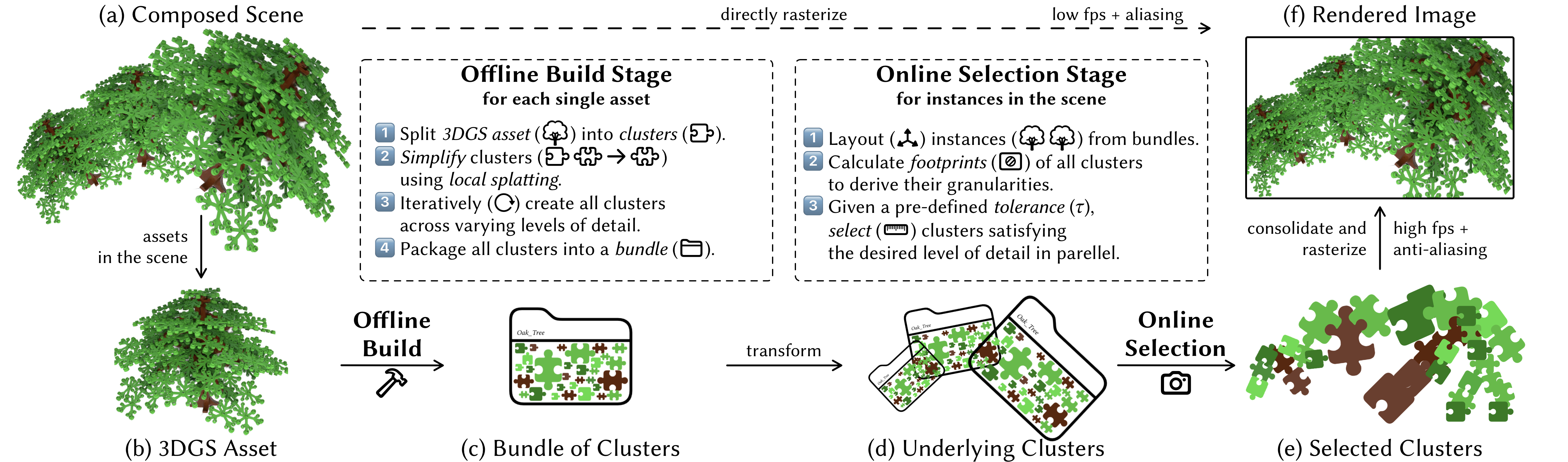}
    \caption{\label{fig:framework}
        \textbf{Framework of our virtualized LOD system}, featuring the \textbf{offline build stage} to prepare clusters at various levels of detail, and the \textbf{online selection stage} to select clusters at appropriate levels of detail given rendering conditions.
    }
    \Description{framework}
\end{figure*}

\subsection{3D Gaussian Splatting for Reconstruction}

Recently, 3D Gaussian Splatting (3DGS)~\cite{gs} has made significant advancements in the Novel View Synthesis (NVS) field by employing explicit 3D Gaussian primitives for reconstruction from captured multi-view images, achieving state-of-the-art rendering quality and speed compared to Neural Radiance Field (NeRF) -based works \cite{nerf,mipnerf360,muller2022instant,fridovich2022plenoxels}.
Each 3D Gaussian primitive is parameterized by a set of attributes, including position $\boldsymbol{p}$, scale $\boldsymbol{s}$, rotation quaternion $\boldsymbol{q}$, opacity $o$, and view-dependent color $\boldsymbol{c}$.
Rendering an image requires first splatting 3D Gaussians onto the image plane \cite{zwicker2001ewa} and then accumulating colors of 2D splats overlapping each pixel using alpha-blending.

In order to capture fine details of objects or scenes being reconstructed, vanilla 3DGS, its variants~\cite{scaffoldgs,mipgs} to better the representation, and some compression works \cite{niemeyer2024radsplat,fan2023lightGaussian,niedermayr2023compressed,papantonakis2024reducing,lee2023compact,navaneet2023compact3d,chen2024hac,yang2024spectrally,zhang2024gaussianspa} to mitigate the high storage requirement, typically employ a high number of 3D Gaussians to derive assets.
These detailed assets captured and reconstructed from the world are ideal for creating a digital world, which may benefit urban planning \cite{proc-gs}, gaming and film production \cite{luma-interactive-scenes}, and XR development \cite{vr-gs}.
However, though the original GPU rasterizer~\cite{gs} can efficiently render each individual asset in real-time, rendering digital worlds with thousands of 3DGS assets in real-time remains a considerable challenge.
The main reason is that, without the control of level of detail, these works regard all 3D Gaussian primitives at the same level of detail, thus presenting over-detailed primitives given low resolution or far camera center. Instead, we feature a cluster-based LOD strategy to selectively render 3D Gaussians at appropriate levels of detail to lower the online rendering overhead.

\subsection{Level-of-Detail for Acceleration}

Level-of-Detail (LOD) techniques are essential in computer graphics for efficiently managing the complexity of 3D assets, enabling real-time rendering applications to display them with minimal computational overhead. 
LOD approaches have been well-established for various types of representations in computer graphics, including meshes \cite{eck1995multiresolution,Stochastic-simplification-of-aggregate-detail,hoppe2023progressive,nanite}, textures \cite{mipmap,ue4_virtual_texturing}, points \cite{QSplat,gobbetti2004layered,wimmer2006instant,wand2007interactive,dachsbacher2003sequential}, and neural representations \cite{mipnerf,mipnerf360,bungeenerf,vrnerf}.
These methods adjust the level of detail based on specific conditions to enhance rendering performance and visual quality.
Notably, appearance-preserving pre-filtering methods \cite{lod-raytracing-bako,lod-raytracing-weier,n-bvh-weier,LEAN-mapping, Linear-efficient-antialiased-displacement-and-reflectance-mapping, Downsampling-scattering-parameters-for-rendering-anisotropic-media, Hybrid-mesh-volume-LoDs-for-all-scale-pre-filtering-of-complex-3D-assets, A-non-exponential-transmittance-model-for-volumetric-scene-representations} assist LOD techniques in generating multiple levels of detail. Similarly, we employ this approach to simplify 3D Gaussians while preserving their visual consistency.

Recently, LOD techniques have been applied to 3D Gaussian representation to efficiently reconstruct and render large scenes \cite{octree-gs,hierarchical-gs,letsgo,FLoD}.
The commonality between our approach and previous work lies in the offline generation of hierarchical structures to accelerate the online rendering of 3DGS assets.
However, a key distinction is that existing methods focus primarily on the NVS task, treating each large scene as an integral and fixed entity.
In contrast, our work extends beyond the NVS context, aiming to accelerate the rendering of complex scenes in digital film and game production, handling multiple high-quality interactive 3D assets with arbitrary positions, rotations, and scales during real-time rendering.
While existing methods, such as Hierarchical 3D Gaussians (H3DG)~\cite{hierarchical-gs}, could be adapted for this scenario, we find that our cluster-based LOD system yields more satisfying rendering results. Additionally, our approach is more general, as it can be applied to both object-level and scene-level 3DGS assets without the need for supervision by captured multi-view images. This extends the 3D Gaussian representation with a robust LOD feature, offering greater flexibility and scalability in real-time applications.

\subsection{Nanite Virtualized Geometry}

Nanite~\cite{nanite} Virtualized Geometry, one of the advanced industrial LOD solutions for meshes and triangle primitives, involves organizing clusters of primitives into multiple layers with different levels of detail, and then selecting only the necessary clusters as the geometry of the scene online to reduce the rendering overhead, which enables the real-time rendering of complex scenes containing billions of primitives while preserving the highly detailed geometry.

Nanite's approach aligns closely with our envisioned LOD system for 3D Gaussians.
The challenge it addresses is similar to ours, and Nanite’s method for managing triangle primitives and geometry is highly adaptable to 3D Gaussian primitives and appearance, as both types of primitives are explicit and have spatial extents. Inspired by this, we design our LOD system for 3D Gaussians.

\section{METHOD\label{sec:method}}

To speed up the rendering of digital worlds with multiple 3DGS assets while maintaining high quality, we propose a Level of Detail (LOD) system, as shown in Fig.~\ref{fig:framework}. This system has two key stages: the \textit{offline build stage} and the \textit{online selection stage}.

\paragraph{Offline Build Stage}
In the offline build stage, we process each 3DGS asset into clusters at different detail levels. Starting with a high-quality 3DGS asset containing millions of 3D Gaussians $G$, we first split these 3D Gaussians into clusters $C_0$ at the finest level (Sec.~\ref{sec:method-clustering}). Then, coarser levels are generated iteratively using a \textit{local splatting} method (Sec.~\ref{sec:method-simplification}). Once all levels are created, they are stored as a bundle for use in the online selection stage. The detailed procedure is described in Algorithm~\ref{alg:lod-build-stage} in the Appendices.

\paragraph{Online Selection Stage}
Using the pre-computed bundles, the online selection stage picks specific clusters to reduce rendering workload while keeping the visual quality high. For this, we use the concept of \textit{footprint} $F$ to measure the detail level needed for each cluster (Sec.~\ref{sec:method-footprint}). Based on a given \textit{footprint tolerance} $\tau$, the system selects appropriate clusters in parallel (Sec.~\ref{sec:method-selection}) and then rasterizes them with the original splatting algorithm~\cite{gs}. This approach reduces the number of 3D Gaussians processed, enabling real-time rendering of complex digital worlds.


\subsection{Clustering Adjacent 3D Gaussians\label{sec:method-clustering}}

\begin{figure}[t]
    \centering
    \includegraphics[width=\linewidth]{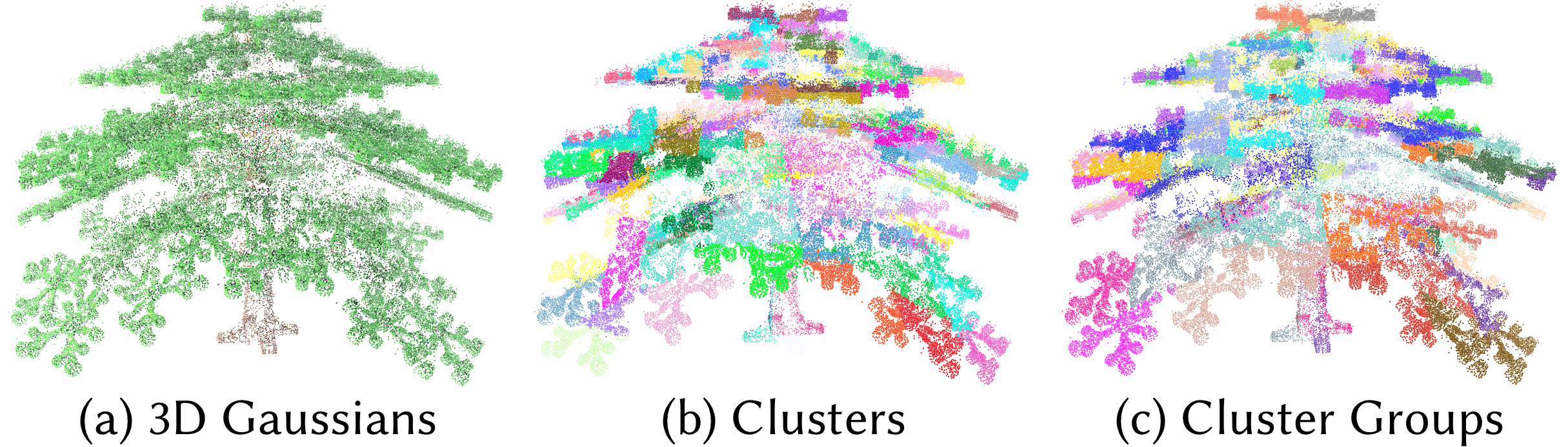}
    \caption{\label{fig:method-clustering}
        \textbf{Clustering on the 3DGS asset} \textsc{oak} from RTMV dataset~\cite{RTMV}.
        (a) \textbf{3D Gaussian primitives} $G$ visualized as colored points.
        (b) \textbf{Clusters} $C_0$ in the finest layer.
        (c) \textbf{Cluster groups} $CG_0$ used for simplification from the finest layer.}
    \Description{method-clustering}
\end{figure}

\begin{figure}[t]
    \centering
    \includegraphics[width=\linewidth]{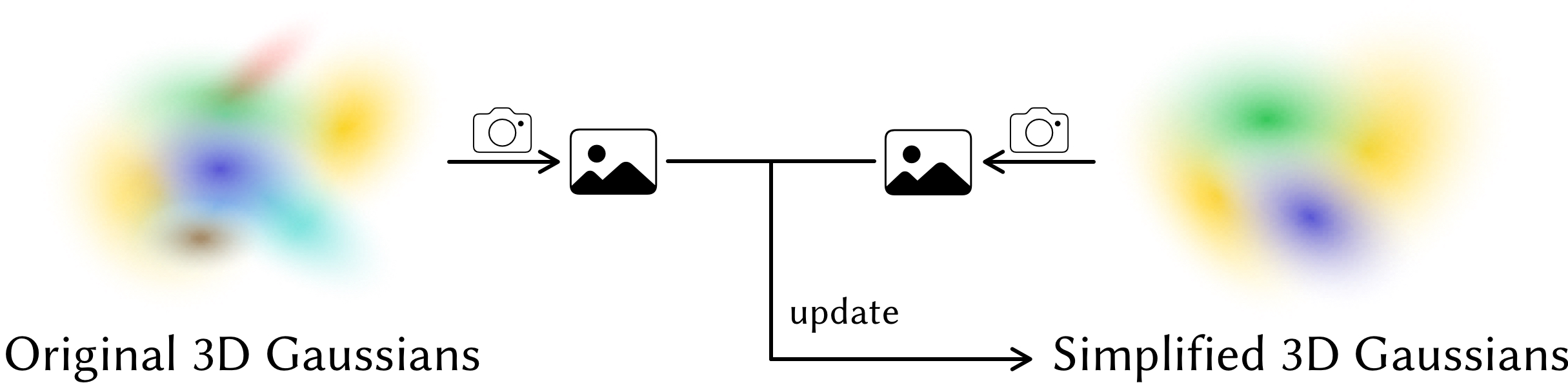}
    \caption{\label{fig:method-local-splatting}
        \textbf{Local splatting method.}
        Simplified 3D Gaussians are optimized locally by distilling the appearance of original 3D Gaussians.
    }
    \Description{method-local-splatting}
\end{figure}

The fundamental unit of our LOD system is the \textit{cluster}, which consists of $n_{G \in C}$ adjacent 3D Gaussians.
Instead of processing $N_G$ 3D Gaussians directly in each 3DGS asset, we first \textit{divide} them into $N_G / n_{G \in C}$ small clusters $C_0$, using their spatial positions as input for the clustering algorithm.
The clustering process is shown in Fig.~\ref{fig:method-clustering} (a)(b). We choose $n_{G \in C} = 4096$, and use the binary median split for clustering as in H3DG~\cite{hierarchical-gs} in our experiments.

Starting from the finest layer with clusters $C_0$, we iteratively \textit{simplify} the clusters, reducing their number by half at each step.
This process continues until the coarsest layer contains only a few clusters, as illustrated in Fig.~\ref{fig:method-tree-structure}.
The details of the \textit{simplification} process are provided in Sec.~\ref{sec:method-simplification}.
The resulting \textit{bundle} comprises multiple clusters across layers with varying levels of detail.
During the online selection stage, appropriate clusters are chosen to meet the required level of detail while minimizing rendering overhead.


\subsection{Simplification in Each Cluster Group \label{sec:method-simplification}}

\begin{figure}[t]
    \centering
    \includegraphics[width=\linewidth]{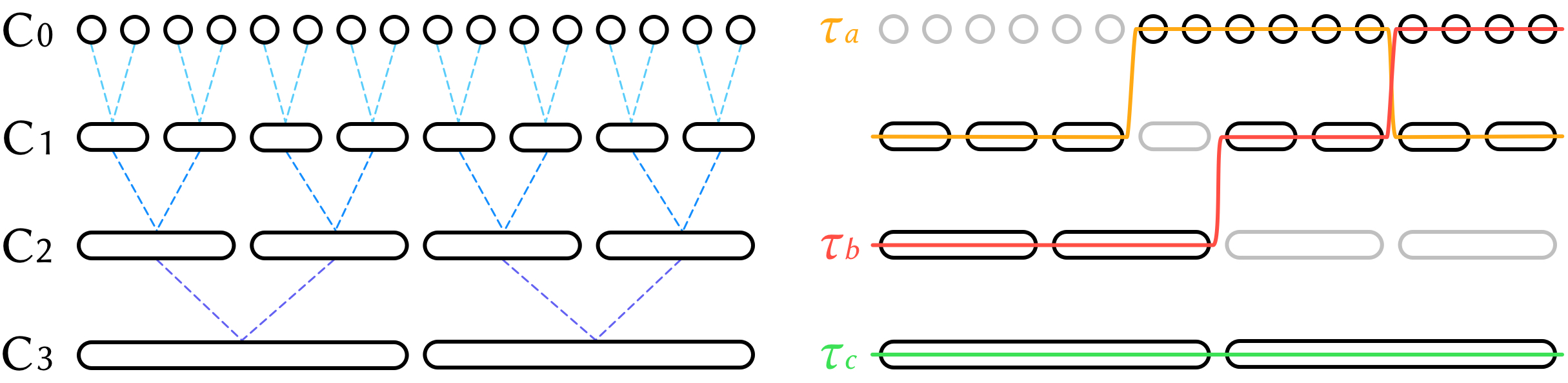}
    \caption{\label{fig:method-tree-structure}
        \textit{Left}: \textbf{Iterative simplification.} Each two adjacent clusters are simplified to create a new cluster, forming the tree structure.
        \textit{Right}: \textbf{Online selection.} Given footprint tolerance $\tau$, clusters are selected with appropriate levels of detail separately.
    }
    \Description{method-tree-structure}
\end{figure}

To halve the number of clusters, we reapply the clustering algorithm, \textit{grouping} $n_{C \in CG}$ adjacent clusters into a \textit{cluster group}, as illustrated in Fig.~\ref{fig:method-clustering} (c).
Within each cluster group, $n_{C \in CG}$ clusters are simplified into $n_{C \in CG} / 2$ clusters. 
While Nanite~\cite{nanite} employs a directed acyclic graph (DAG) structure with $n_{C \in CG} > 2$ to mitigate the appearance of dense cruft, we simply adopt a \textit{tree} structure with $n_{C \in CG} = 2$ (Fig.~\ref{fig:method-tree-structure}). We find this approach effective, owing to the inherently unstructured nature of 3D Gaussians (see Appendix~\ref{sec:appendices-ablation-dag-and-tree} for details).

The simplification begins by extracting the \textit{original 3D Gaussians} from each pair of adjacent clusters and downsampling them by half, retaining those with larger products of scale and opacity, as these factors have a greater impact on appearance. Their scales are then slightly increased by a factor of $2^{1/6}$ to form the \textit{initial simplified 3D Gaussians}.
Subsequently, similar to the distillation process, we \textit{optimize} the simplified 3D Gaussian to preserve as much visual detail as possible.
Finally, the optimized 3D Gaussians are consolidated into a single cluster, the basic unit in our system.

To preserve visual quality while reducing the number of 3D Gaussians by half, we develop the \textit{local splatting} method, leveraging the differentiable properties of 3D Gaussians (Fig.~\ref{fig:method-local-splatting}).
A total of 640 randomly generated pseudo-views are positioned to face the center of the cluster group, each placed at a distance equal to four times the cluster group’s radius. These views are used to render both the original and simplified 3D Gaussians at a resolution of 64x64. A loss is calculated by comparing the two renders, which is then back-propagated to optimize the properties of the simplified 3D Gaussians. During iterative optimization, different from 3DGS~\cite{gs}, we disable adaptive control of 3D Gaussians, employ a lower learning rate (1.6e-5) for position updates, and introduce alpha channel supervision (see Appendix~\ref{sec:appendices-datasets}) to mitigate background color influence. This process ensures that spatial coherence and visual quality are maintained within the cluster group after simplification.


\subsection{Footprint Analysis of Cluster Group \label{sec:method-footprint}}

\begin{figure}[t]
    \centering

    \begin{tikzpicture}
    \node[anchor=south west] (image) at (0, 0.03\linewidth) 
        {
        \includegraphics[width=0.5\linewidth]{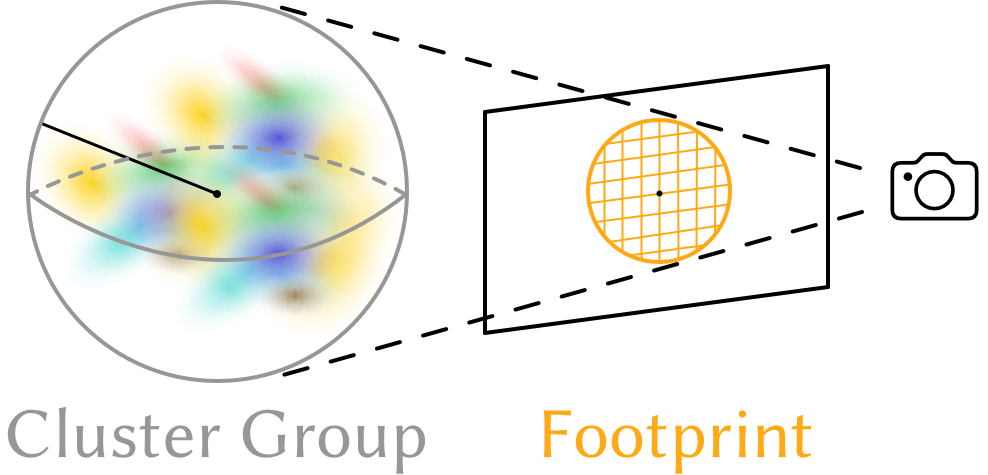}
        };
    \node[anchor=south west] at (0.56\linewidth, 0.0\linewidth)
        {$
        \begin{aligned}
        \left\{
            \begin{aligned}
            \boldsymbol{c}_w &= \frac{1}{N_G} \sum_{G} \boldsymbol{p}_G\\
            r &= \max_{G}{\left\lVert \boldsymbol{p}_{G} - \boldsymbol{c}_w \right\rVert}
            \end{aligned}
        \right.\\
        F = \frac{a}{4} w h = \pi \frac{r^2}{z_v^2} f_x f_y
        \end{aligned}
        $};
    \node[anchor=south west] at (0.06\linewidth, 0.20\linewidth)
        {$r$};
    \node[anchor=south west] at (0.12\linewidth, 0.16\linewidth)
        {$\boldsymbol{c}_w$};
    \node[anchor=south west] at (0.27\linewidth, 0.16\linewidth)
        {\textcolor{my-color-footprint-orange}{$F$}};
    \node[anchor=south west] at (0.375\linewidth, 0.16\linewidth)
        {$\boldsymbol{c}_v$};
    \end{tikzpicture}
    
    \caption{\label{fig:method-footprint}
        \textbf{Footprint of a cluster group.}
        The bounding sphere of a cluster group is projected onto screen space, occupying several pixels as footprint.
    }
    \Description{method-footprint}
\end{figure}
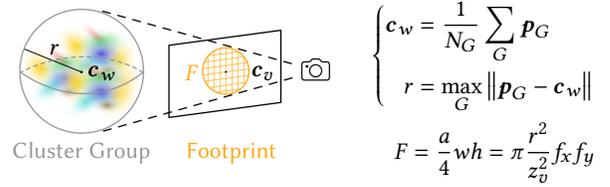

\begin{figure*}[t]
    \centering
    \includegraphics[width=\linewidth]{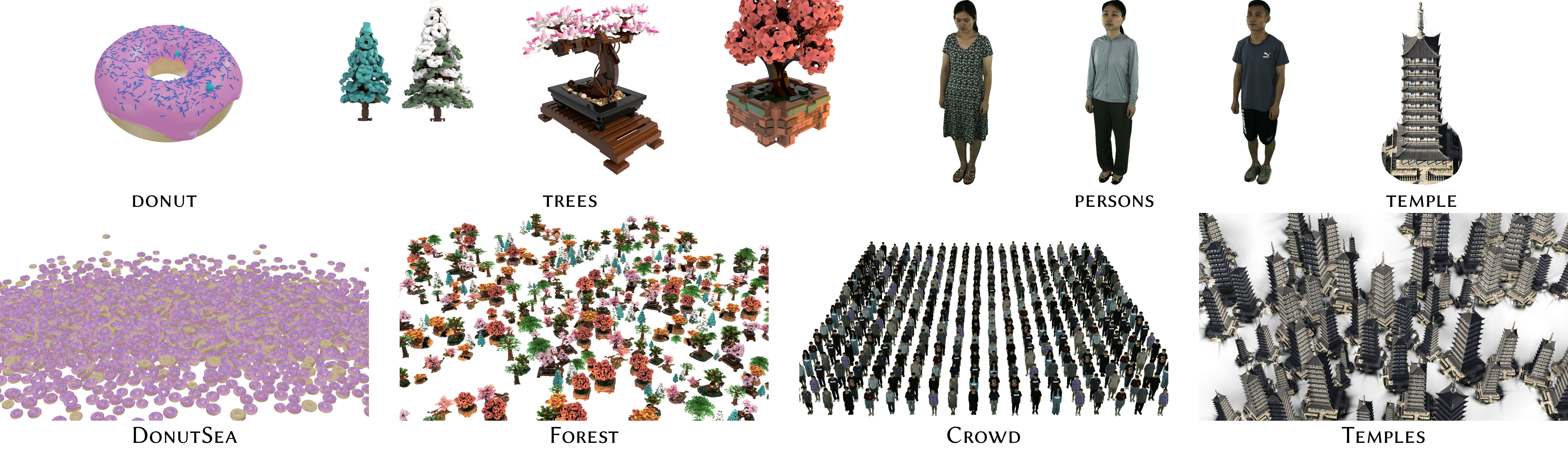}
    \caption{\label{fig:dataset-consolidated}
        \textbf{Representative objects and composed scenes} from daily life in our curated dataset.
        \textit{Top}: Individual 3D Gaussian objects spanning categories such as food, trees, humans, and buildings.
        \textit{Bottom}: Various composed scenes with 3DGS assets, each containing around 0.1 billion of 3D Gaussians.
    }
    \Description{dataset-consolidated}
\end{figure*}

During the simplification, "coarser" clusters are derived from "finer" clusters.
To enable selection of clusters at varying levels of detail using a single criterion, we quantify the "level of detail" using the \textit{footprint} of each cluster group, as shown in Fig.~\ref{fig:method-footprint}.

A cluster group is modeled as a bounding sphere, with its center $\boldsymbol{c}_w = (x_w, y_w, z_w)$ in the world space computed as the mean position of 3D Gaussians it contains. The radius $r$ is the farthest distance from the center to any 3D Gaussians in the cluster group.
In the camera space, the sphere's center transforms to $\boldsymbol{c}_v = (x_v, y_v, z_v)$, and its projection onto the screen forms an ellipse.
The approximated ellipse's area $a$ in normalized device coordinates defines the \textit{footprint} $F$, which represents the number of pixels occupied by the cluster group on a screen with resolution $w \times h$ (where $f_x$ and $f_y$ are the camera focal lengths).

The footprint generally represents the level of detail: larger footprints indicate lower detail, since each cluster group contains a similar number of 3D Gaussians.
As the footprint varies with camera perspective, it cannot be precomputed during the offline stage. Instead, the centers and radii of all cluster groups are pre-calculated and stored for real-time rendering. Using the concept of footprint, we detail the selection process in Sec.~\ref{sec:method-selection}.


\subsection{Parallel Selection on Discrete Clusters \label{sec:method-selection}}

As explained in Sec.~\ref{sec:method-footprint}, when a cluster group is simplified, its center $\boldsymbol{c}$ and radius $r$ are computed, and cluster groups are selected based on their screen space footprint $F$.
However, cluster groups serve only as intermediate structures for simplification and footprint calculation. In our LOD system, the smallest unit is always the cluster.

To prevent performance issues from traversing the entire tree of clusters during rendering, we implement parallel cluster selection.
During the build stage, the center $\boldsymbol{c}$ and radius $r$ of each cluster group are assigned to the clusters it contains.
Each simplified cluster (child) stores both its own spatial data ($\boldsymbol{c}_c$, $r_c$) and the corresponding parent cluster's data ($\boldsymbol{c}_p$ and $r_p$).
We ensure $r_p > r_c$, maintaining a monotonic tree structure and preventing simultaneous selection of clusters that overlap in 3D space.

During the selection stage, the pre-stored spatial data pairs $(\boldsymbol{c}_c, r_c)$ and $(\boldsymbol{c}_p, r_p)$ are retrieved for each cluster.
Using the camera view $V$ and image resolution $w \times h$, footprints $F_c$ and $F_p$ are calculated in parallel.
These footprints are then compared against a predefined \textit{footprint tolerance} $\tau$ to decide whether to render a cluster. Only clusters satisfying $\boldsymbol{F_c \leq \tau < F_p}$ are rendered.
This approach ensures that the selected clusters meet the desired level of detail, while their parent clusters fail the tolerance check, achieving optimal LOD selection.
An example of the selection process with three different $\tau$s is illustrated in Fig.~\ref{fig:method-tree-structure}. 
A key advantage of our approach is that the tolerance can be user-defined during the online rendering process, allowing for a flexible trade-off between performance and visual quality. The specific tolerances used for different scenes in our experiments are indicated in Fig.~\ref{fig:teaser}, Fig.~\ref{fig:experiment-visual}, and Fig.~\ref{fig:ablation-visual}.

\section{Dataset}

To evaluate the effectiveness of the proposed system, we create a dataset comprising several 3DGS assets reconstructed using the 3DGS algorithm (Sec.~\ref{sec:datasets-3dgs-assets}) and organize them into four composed scenes (Sec.~\ref{sec:datasets-composed-scenes}), representing a digital world proxy.

\subsection{3DGS Assets\label{sec:datasets-3dgs-assets}}

As shown in Fig.~\ref{fig:dataset-consolidated}, collected assets are diverse and include:
(synthetic) A \textsc{donut} sourced from the Internet~\cite{donut-asset};
Eight \textsc{trees} from the Bricks environment in the RTMV dataset~\cite{RTMV};
(real) Sixteen \textsc{persons} from the MVHumanNet dataset~\cite{mvhumannet};
A \textsc{temple} captured by us.
These 3DGS assets are reconstructed from multi-view images using Mip-Splatting~\cite{mipgs}, a variant of the vanilla 3DGS algorithm.
Details on the reconstruction process and analysis of assets are provided in Appendix~\ref{sec:appendices-datasets-acquisition-of-3dgs-assets}.

\subsection{Composed Scenes\label{sec:datasets-composed-scenes}}

We further construct composed scenes from these assets, by arranging them in a blank environment, as illustrated in Fig.~\ref{fig:dataset-consolidated}.
These scenes include:
\textsc{DonutSea} simulating fallen \textsc{donut}s on the ground using Blender;
\textsc{Crowd} placing \textsc{persons} in a grid pattern;
\textsc{Forest} and \textsc{Temples} arranging \textsc{trees} and \textsc{temple}s on the ground with random 2D rotations and uniform scales.
Each scene comprises approximately 100 million 3D Gaussians, presenting a significant challenge for real-time rendering when most 3D Gaussians appear within the camera frustum.
Details on the construction of these composed scenes are provided in Appendix~\ref{sec:appendices-datasets-construction-of-composed-scenes}.

\section{EXPERIMENTS\label{sec:experiments}}


\subsection{Implementation Details\label{sec:experiments-implementation-details}}

We use the gsplat library~\cite{gsplat} as the rasterizer for 3D Gaussians, rendering at a resolution of 1920x1080 with a field of view of $fov_x = \pi / 4$.
As in the asset acquisition process (Appendix~\ref{sec:appendices-datasets-acquisition-of-3dgs-assets}), we remove the dilation operator in vanilla 3DGS to address the dilation issue by setting \textit{eps2d} to 0.0 following Mip-Splatting~\cite{mipgs}.
All our experiments are conducted on a single NVIDIA A800 GPU with 80 GB memory.
The source code and the dataset are available at \url{https://github.com/city-super/V3DG}.


\subsection{Metrics\label{sec:experiments-metrics}}

We conduct comprehensive evaluations on the proposed dataset.
To evaluate our method, we design the following camera trajectories.
From the center of each scene, we choose four orthogonal directions on the $xy$-plane, with five elevation angles of 15°, 30°, 45°, 60°, and 75°. Along each direction, 20 cameras centered at the coordinate origin are evenly spaced within the scene's extent, deriving 20 relative distances for each scene.

We adopt three metrics to quantitatively assess the performance for rendering of our system compared to vanilla 3DGS: (1) \textit{number and percentage of selected 3D Gaussians} to show our system selecting only a subset of 3D Gaussians, (2) \textit{FPS and acceleration rate} to represent the multiple of speed our system achieves, and (3) \textit{FLIP error}~\cite{flip} to evaluate the human-perceived differences when alternating between renders of our system and vanilla 3DGS. The lower the FLIP error, the less difference can be perceived.
Notably, we observe that vanilla 3DGS renders at 1080p suffer from significant aliasing, making them unsuitable for direct comparison. To address this, we render 4K images using vanilla 3DGS and apply SuperSampling Anti-Aliasing (SSAA) to obtain high-quality, anti-aliased 1080p reference images (denoted as 3DGS-SSAA). We then compare both vanilla 3DGS (3DGS) and our method (Ours), rendered at 1080p, against these SSAA-derived "ground truth" images.


\subsection{Results\label{sec:experiments-results}}

\begin{figure*}[t]
    \centering
    \includegraphics[width=0.96\linewidth]{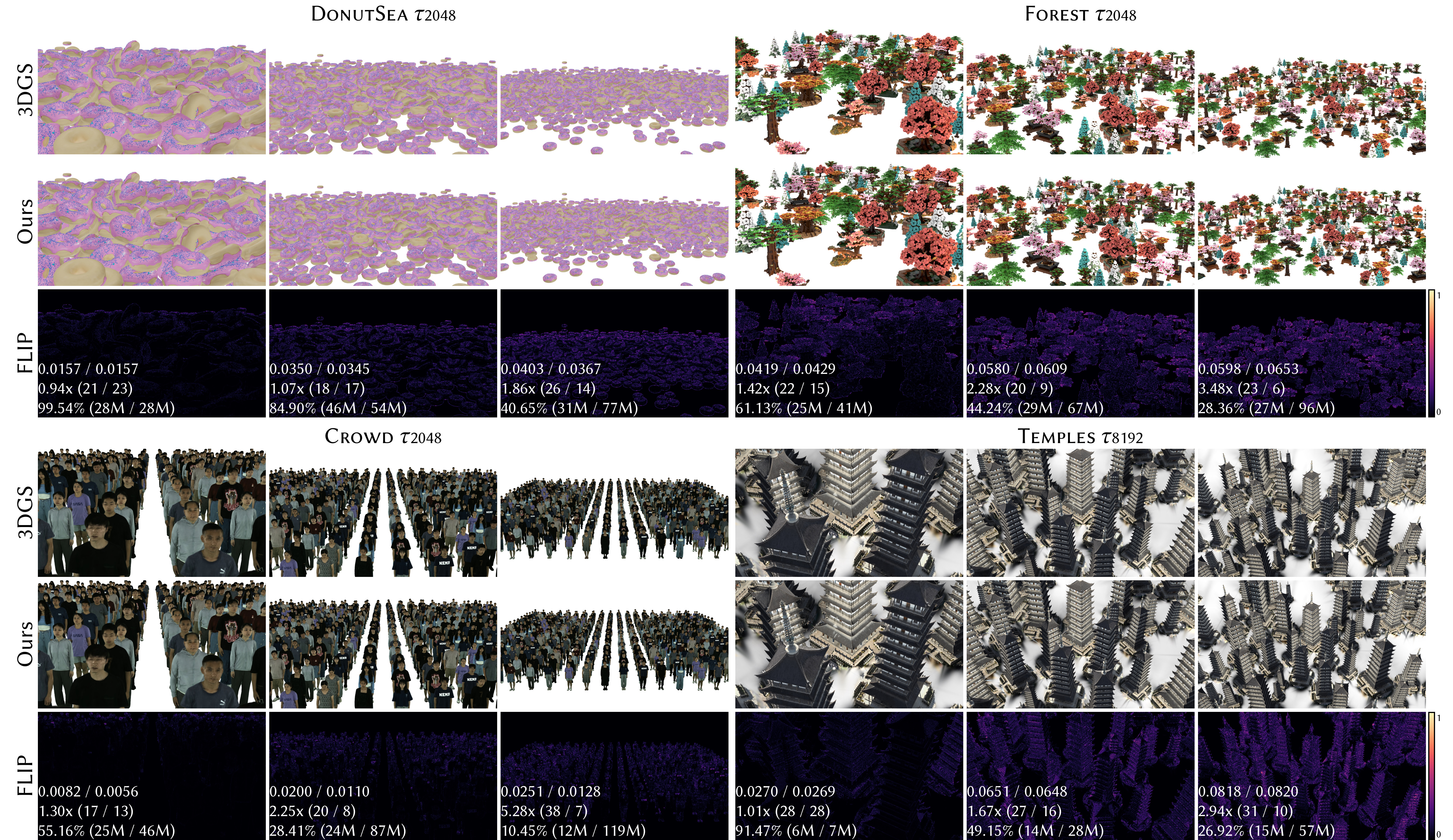}

    \vspace{-6pt}
    
    \caption{\label{fig:experiment-visual}
        \textbf{Renders on composed scenes.}
        We show renders of 3DGS and our system across four composed scenes at three different distances.
        FLIP errors, acceleration rates (x), and percentages of selected Gaussians (\%) are indicated at the bottom-left corner (Ours / 3DGS).
    }
    \Description{experiment-visual}
\end{figure*}

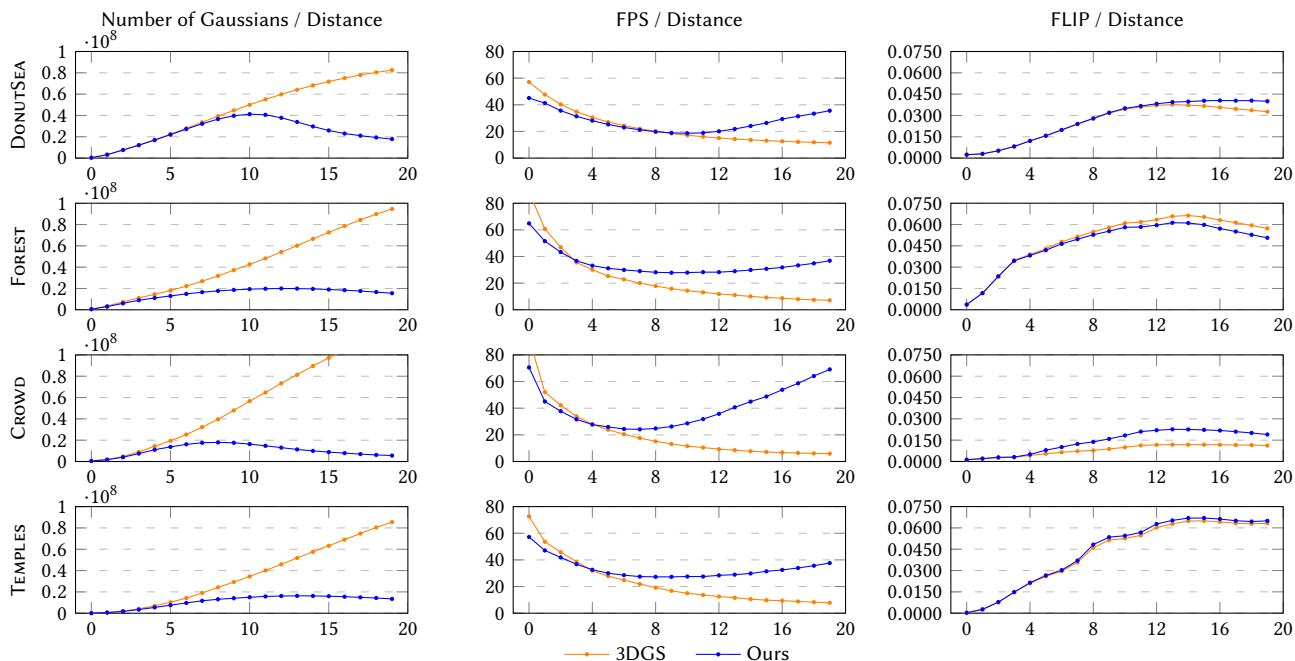
\begin{figure*}[t]

\pgfplotstableread[col sep=comma]{graphs/data/data-Forest.csv}\DataForest
\pgfplotstableread[col sep=comma]{graphs/data/data-DonutSea.csv}\DataDonutSea
\pgfplotstableread[col sep=comma]{graphs/data/data-Crowd.csv}\DataCrowd
\pgfplotstableread[col sep=comma]{graphs/data/data-Temples.csv}\DataTemples

\begin{tikzpicture}[font=\small\sffamily]
    \begin{groupplot}[
        group style = {
            group size = 3 by 4,
            horizontal sep = 1.4cm,
            vertical sep = 0.6cm,
        },
        height=3.0cm,
        width=6cm
    ]

    
    \nextgroupplot[
        title = Number of Gaussians / Distance,
        ylabel = \textsc{DonutSea},
        xmin = -1.0,
        xmax = 20.0,
        ymin = 0,
        ymax = 100000000,
        ymajorgrids = true,
        grid style=loosely dashed,
    ]
    \addplot[color=orange, mark=*, mark size=0.6pt]
    table[x={distance}, y={gs_count}] from \DataDonutSea;
    \addplot[color=blue, mark=*, mark size=0.6pt]
    table[x={distance}, y={vg_count}] from \DataDonutSea;


    \nextgroupplot[
        title = FPS / Distance,
        xmin = -1.0,
        xmax = 20.0,
        xtick distance = 4,
        ymin = 0,
        ymax = 80,
        ymajorgrids=true,
        grid style=loosely dashed,
    ]
    \addplot[color=orange, mark=*, mark size=0.6pt]
    table[x={distance}, y={gs_fps}] from \DataDonutSea;
    \addplot[color=blue, mark=*, mark size=0.6pt]
    table[x={distance}, y={vg_fps}] from \DataDonutSea;


    \nextgroupplot[
        title = FLIP / Distance,
        xmin = -1.0,
        xmax = 20.0,
        xtick distance = 4,
        ymin = 0,
        ymax = 0.0750,
        ytick distance = 0.0150,
        scaled ticks=false,
        y tick label style={/pgf/number format/fixed, /pgf/number format/precision=4, /pgf/number format/fixed zerofill},
        ymajorgrids=true,
        grid style=loosely dashed,
    ]
    \addplot[color=orange, mark=*, mark size=0.6pt]
    table[x={distance}, y={gs_flip}] from \DataDonutSea;
    \addplot[color=blue, mark=*, mark size=0.6pt]
    table[x={distance}, y={vg_flip}] from \DataDonutSea;


    \nextgroupplot[
        ylabel = \textsc{Forest},
        xmin = -1.0,
        xmax = 20.0,
        ymin = 0,
        ymax = 100000000,
        ymajorgrids = true,
        grid style=loosely dashed,
    ]
    \addplot[color=orange, mark=*, mark size=0.6pt]
    table[x={distance}, y={gs_count}] from \DataForest;
    \addplot[color=blue, mark=*, mark size=0.6pt]
    table[x={distance}, y={vg_count}] from \DataForest;

    
    \nextgroupplot[
        xmin = -1.0,
        xmax = 20.0,
        xtick distance = 4,
        ymin = 0,
        ymax = 80,
        ymajorgrids=true,
        grid style=loosely dashed,
    ]
    \addplot[color=orange, mark=*, mark size=0.6pt]
    table[x={distance}, y={gs_fps}] from \DataForest;
    \addplot[color=blue, mark=*, mark size=0.6pt]
    table[x={distance}, y={vg_fps}] from \DataForest;

    
    \nextgroupplot[
        xmin = -1.0,
        xmax = 20.0,
        xtick distance = 4,
        ymin = 0,
        ymax = 0.0750,
        ytick distance = 0.0150,
        scaled ticks=false,
        y tick label style={/pgf/number format/fixed, /pgf/number format/precision=4, /pgf/number format/fixed zerofill},
        ymajorgrids=true,
        grid style=loosely dashed,
    ]
    \addplot[color=orange, mark=*, mark size=0.6pt]
    table[x={distance}, y={gs_flip}] from \DataForest;
    \addplot[color=blue, mark=*, mark size=0.6pt]
    table[x={distance}, y={vg_flip}] from \DataForest;


    \nextgroupplot[
        ylabel = \textsc{Crowd},
        xmin = -1.0,
        xmax = 20.0,
        ymin = 0,
        ymax = 100000000,
        ymajorgrids = true,
        grid style=loosely dashed,
    ]
    \addplot[color=orange, mark=*, mark size=0.6pt]
    table[x={distance}, y={gs_count}] from \DataCrowd;
    \addplot[color=blue, mark=*, mark size=0.6pt]
    table[x={distance}, y={vg_count}] from \DataCrowd;

    
    \nextgroupplot[
        xmin = -1.0,
        xmax = 20.0,
        xtick distance = 4,
        ymin = 0,
        ymax = 80,
        ymajorgrids=true,
        grid style=loosely dashed,
    ]
    \addplot[color=orange, mark=*, mark size=0.6pt]
    table[x={distance}, y={gs_fps}] from \DataCrowd;
    \addplot[color=blue, mark=*, mark size=0.6pt]
    table[x={distance}, y={vg_fps}] from \DataCrowd;
    
    
    \nextgroupplot[
        xmin = -1.0,
        xmax = 20.0,
        xtick distance = 4,
        ymin = 0,
        ymax = 0.0750,
        ytick distance = 0.0150,
        scaled ticks=false,
        y tick label style={/pgf/number format/fixed, /pgf/number format/precision=4, /pgf/number format/fixed zerofill},
        ymajorgrids=true,
        grid style=loosely dashed,
    ]
    \addplot[color=orange, mark=*, mark size=0.6pt]
    table[x={distance}, y={gs_flip}] from \DataCrowd;
    \addplot[color=blue, mark=*, mark size=0.6pt]
    table[x={distance}, y={vg_flip}] from \DataCrowd;


    \nextgroupplot[
        ylabel = \textsc{Temples},
        xmin = -1.0,
        xmax = 20.0,
        ymin = 0,
        ymax = 100000000,
        ymajorgrids = true,
        grid style=loosely dashed,
    ]
    \addplot[color=orange, mark=*, mark size=0.6pt]
    table[x={distance}, y={gs_count}] from \DataTemples;
    \addplot[color=blue, mark=*, mark size=0.6pt]
    table[x={distance}, y={vg_count}] from \DataTemples;

    
    \nextgroupplot[
        xmin = -1.0,
        xmax = 20.0,
        xtick distance = 4,
        ymin = 0,
        ymax = 80,
        ymajorgrids=true,
        grid style=loosely dashed,
        legend style={
            draw=none,
            anchor=north,   
            at={(0.5,-0.20)},
            legend columns=-1,
            /tikz/every even column/.append style={column sep=1.0em},
        },
    ]
    \addplot[color=orange, mark=*, mark size=0.6pt]
    table[x={distance}, y={gs_fps}] from \DataTemples;
    \addlegendentry{3DGS}
    \addplot[color=blue, mark=*, mark size=0.6pt]
    table[x={distance}, y={vg_fps}] from \DataTemples;
    \addlegendentry{Ours}

    
    \nextgroupplot[
        xmin = -1.0,
        xmax = 20.0,
        xtick distance = 4,
        ymin = 0,
        ymax = 0.0750,
        ytick distance = 0.0150,
        scaled ticks=false,
        y tick label style={/pgf/number format/fixed, /pgf/number format/precision=4, /pgf/number format/fixed zerofill},
        ymajorgrids=true,
        grid style=loosely dashed,
    ]
    \addplot[color=orange, mark=*, mark size=0.6pt]
    table[x={distance}, y={gs_flip}] from \DataTemples;
    \addplot[color=blue, mark=*, mark size=0.6pt]
    table[x={distance}, y={vg_flip}] from \DataTemples;

    \end{groupplot}
\end{tikzpicture}

\vspace{-14pt}

\caption{\label{fig:experiment-composed-scenes-graph}
    \textbf{Graphs on composed scenes.}
    We report metrics comparing renders of 3DGS and our system across four composed scenes at twenty relative distances.
}
\Description{experiment-composed-scenes-graph}
\end{figure*}

The qualitative and quantitative results are presented in Fig.~\ref{fig:experiment-visual} and Fig.~\ref{fig:experiment-composed-scenes-graph} on four composed scenes.

As the viewing distance from the center of the scene increases, although more 3D Gaussians enter the camera frustum, our system selectively renders the necessary clusters with appropriate levels of detail.
For instance, in Fig.~\ref{fig:experiment-visual}, when viewing the \textsc{Forest} from the "near", "middle", and "far" distances, 3DGS needs to render 41M, 67M, and 96M 3D Gaussians, while our system selectively renders 25M, 29M, and 27M 3D Gaussians, as utilizing excessive 3D Gaussians to depict insignificant details that exceed the rendering resolution and human visual perception is unnecessary.
The first column in Fig.~\ref{fig:experiment-composed-scenes-graph} also shows our system helps maintain a stable online rendering cost with a given resolution, having the ability of rendering complex composed scenes with numerous assets in real-time.
This strategy effectively limits the number of 3D Gaussians to rasterize and substantially accelerates the rendering process.
Importantly, at near distances, our system selects a similar number of 3D Gaussians compared to vanilla 3DGS, ensuring that details remain visible when zooming in.

As for the acceleration ability, as presented in Fig.~\ref{fig:experiment-visual}, our system achieves acceleration rates of 1.86x, 3.48x, 5.28x, and 2.94x respectively on four composed scenes at the "far distance".
And as shown in the second column in Fig.~\ref{fig:experiment-composed-scenes-graph}, our system achieves an average of 6.19x acceleration rate at the farthest relative distance.

\begin{figure}[t]
    \centering
    \includegraphics[width=\linewidth]{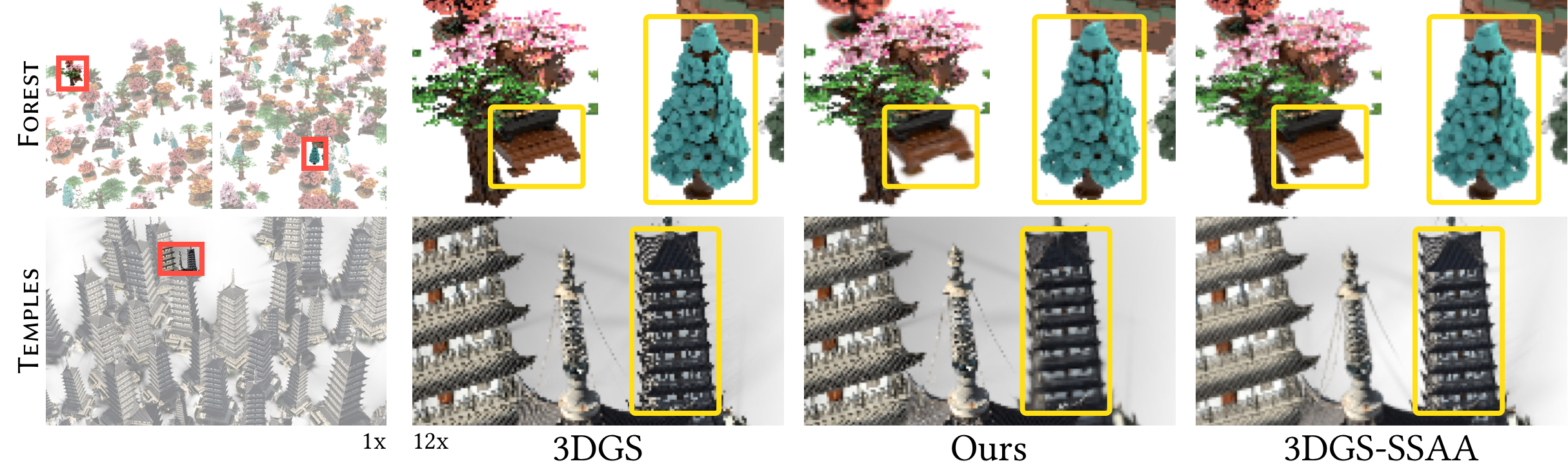}
    \caption{\label{fig:Experiment-anti-aliasing}
        \textbf{Anti-aliasing for visual comfort.}
        Our system \textbf{reduces aliasing effects} by selecting fewer 3D Gaussians, achieving a natural blurring effect when rendering complex scenes with numerous assets.
    }
    \Description{Experiment-anti-aliasing}
\end{figure}

In addition to selection of 3D Gaussians and substantial acceleration, our method achieves visual quality comparable to that of vanilla 3DGS, as evidenced by the closely matching FLIP values shown in Fig.~\ref{fig:experiment-visual} and Fig.~\ref{fig:experiment-composed-scenes-graph}.
Besides, since our system shows appropriate levels of detail matching the rendering resolution, our system also mitigates aliasing issues that can arise from under-sampling large numbers of 3D Gaussians with a limited number of pixels.
We present magnified renders of anti-aliasing effects in Fig.~\ref{fig:Experiment-anti-aliasing}, which demonstrates that our system produces blurring renders that are more visually comfortable, closer to the SSAA renders.

A video roaming in four composed scenes is prepared in the supplementary material to show comparison between renders of our system and 3DGS.
Aliasing and flickering effects are obvious for 3DGS in the video, while frames of our system significantly mitigate such uncomfortable artifacts.
We present some additional analysis on renders of our system in Appendix~\ref{sec:appendix-detailed-analysis-on-renders}.

\begin{table}[t]
    \caption{\label{tab:ablations-mine-forest}
        \textbf{Metrics from ablation studies} on the composed scene \textsc{Forest}, including different simplification iterations and footprint tolerances.
        The basic setting is \underline{underlined} and changing values are \textbf{bold}.
    }
    \begin{tabular}{ rr | cccc }
    \toprule
    ~ & ~ & Duration & Percentage & Rate & FLIP \\
    \hline
    \multicolumn{2}{ c | }{3DGS} & - & 100.00\% & 1.00x & 0.0509 \\
    \hline
    \parbox[t]{2mm}{\multirow{5}{*}{\rotatebox[origin=c]{90}{Iteration}}}
        & $^{\ast}$0   & \textbf{0.1m} & 34.14\% & 1.47x & \textbf{0.0800} \\
        & 0            & \textbf{0.1m} & 33.15\% & 1.45x & \textbf{0.0708} \\
        & 40           & \textbf{0.7s} & 34.20\% & 1.45x & \textbf{0.0522} \\
        & 160          & \textbf{2.5m} & 34.25\% & 1.46x & \textbf{0.0470} \\
        & \underline{640} & \underline{\textbf{9.8m}} & \underline{34.28\%} & \underline{1.47x} & \underline{\textbf{0.0462}} \\
    \hline
    \parbox[t]{2mm}{\multirow{5}{*}{\rotatebox[origin=c]{90}{Tolerance}}}
        & 512           & - & \textbf{72.83\%} & \textbf{1.00x} & \textbf{0.0500} \\
        & 1024          & - & \textbf{53.86\%} & \textbf{1.15x} & \textbf{0.0484} \\
        & \underline{2048} & - & \textbf{\underline{34.28\%}} & \textbf{\underline{1.47x}} & \textbf{\underline{0.0462}} \\
        & 4096          & - & \textbf{18.76\%} & \textbf{2.19x} & \textbf{0.0462} \\
        & 8192          & - & \textbf{9.39\%} & \textbf{3.53x} & \textbf{0.0533} \\
    \bottomrule
    \end{tabular}
\end{table}

\subsection{Ablations\label{sec:ablations}}

\begin{figure*}[t]
    \centering
    \includegraphics[width=\linewidth]{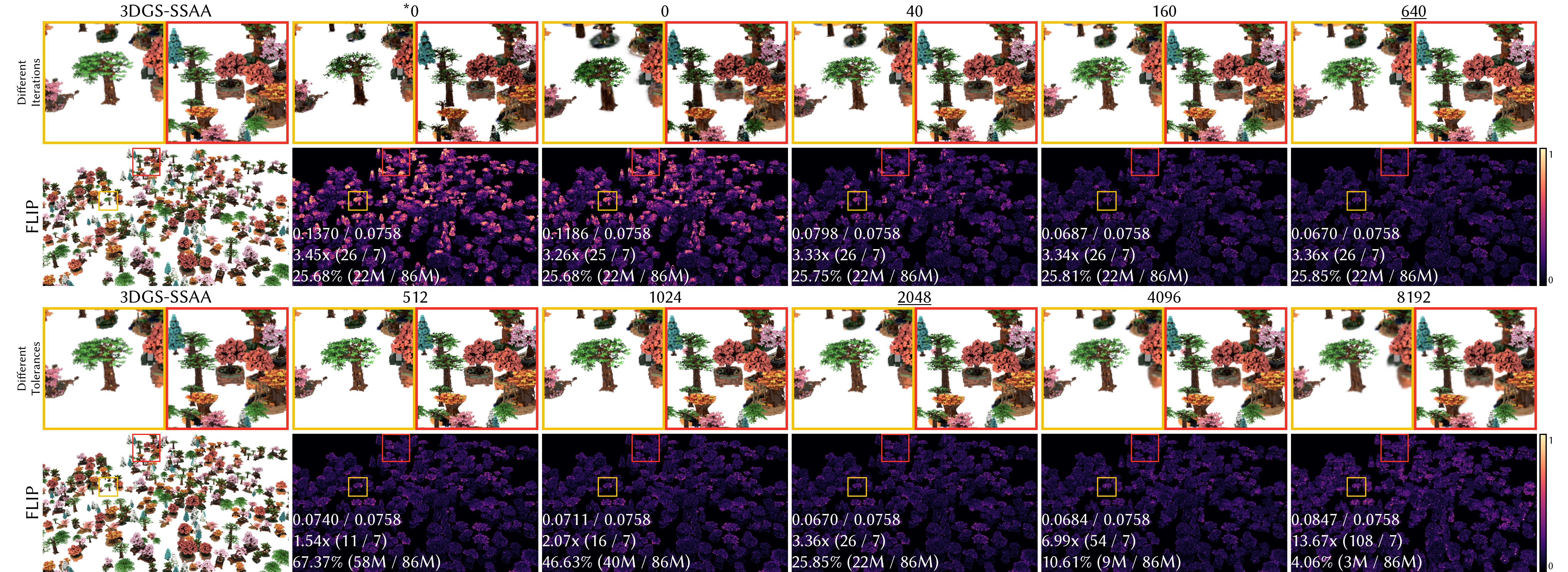}

    \caption{\label{fig:ablation-visual}
        \textbf{Renders from ablation studies.}
        We show renders of our system comparing with 3DGS under varying configurations on the composed scene \textsc{Forest} using the same camera view.
        FLIP errors, acceleration rates (x), and percentages of selected Gaussians (\%) are indicated at the bottom-left corner (Ours / 3DGS).
    }
    \Description{ablation-visual}
\end{figure*}

\begin{figure*}[t]

\pgfplotstableread[col sep=comma]{graphs/data/data-ablations-forest.csv}\DataCSV

\begin{tikzpicture}[font=\small\sffamily]
    \begin{groupplot}[
        group style = {
            group size = 3 by 2,
            horizontal sep = 1.4cm,
        },
        height=3.2cm,
        width=6cm
    ]

    
    \nextgroupplot[
        title = Percentage of Gaussians / Distance,
        align = center,
        ylabel = Different\\Iterations,
        xmin = -1.0,
        xmax = 20.0,
        xtick distance = 4,
        ymin = 0,
        ymax = 100,
        ytick distance = 25,
        ymajorgrids = true,
        grid style=loosely dashed,
        legend style={
            draw=none,
            anchor=north,
            at={(1.8,-0.18)},
            legend columns=5,
            /tikz/every even column/.append style={column sep=1.0em},
        },
    ]
    
    \addplot[color=blue, mark=*, mark size=0.6pt]
    table[x={distance}, y={iter0prime_percentage}] from \DataCSV;
    \addlegendentry{$^\ast$ 0}
    \addplot[color=cyan, mark=*, mark size=0.6pt]
    table[x={distance}, y={iter0_percentage}] from \DataCSV;
    \addlegendentry{0}
    \addplot[color=green, mark=*, mark size=0.6pt]
    table[x={distance}, y={iter40_percentage}] from \DataCSV;
    \addlegendentry{40}
    \addplot[color=orange, mark=*, mark size=0.6pt]
    table[x={distance}, y={iter160_percentage}] from \DataCSV;
    \addlegendentry{160}
    \addplot[color=red, mark=*, mark size=0.6pt]
    table[x={distance}, y={iter640_percentage}] from \DataCSV;
    \addlegendentry{\underline{640}}


    \nextgroupplot[
        title = Acceleration Rate / Distance,
        xmin = -1.0,
        xmax = 20.0,
        xtick distance = 4,
        ymin = 0,
        ymajorgrids=true,
        grid style=loosely dashed,
    ]
    
    \addplot[color=blue, mark=*, mark size=0.6pt]
    table[x={distance}, y={iter0prime_rate}] from \DataCSV;
    \addplot[color=cyan, mark=*, mark size=0.6pt]
    table[x={distance}, y={iter0_rate}] from \DataCSV;
    \addplot[color=green, mark=*, mark size=0.6pt]
    table[x={distance}, y={iter40_rate}] from \DataCSV;
    \addplot[color=orange, mark=*, mark size=0.6pt]
    table[x={distance}, y={iter160_rate}] from \DataCSV;
    \addplot[color=red, mark=*, mark size=0.6pt]
    table[x={distance}, y={iter640_rate}] from \DataCSV;

    
    \nextgroupplot[
        title = FLIP / Distance,
        xmin = -1.0,
        xmax = 20.0,
        xtick distance = 4,
        ymin = 0,
        ymax = 0.120,
        ytick distance = 0.040,
        scaled ticks=false,
        y tick label style={/pgf/number format/fixed, /pgf/number format/precision=4, /pgf/number format/fixed zerofill},
        ymajorgrids=true,
        grid style=loosely dashed,
    ]
    
    \addplot[color=blue, mark=*, mark size=0.6pt]
    table[x={distance}, y={iter0prime_flip}] from \DataCSV;
    \addplot[color=cyan, mark=*, mark size=0.6pt]
    table[x={distance}, y={iter0_flip}] from \DataCSV;
    \addplot[color=green, mark=*, mark size=0.6pt]
    table[x={distance}, y={iter40_flip}] from \DataCSV;
    \addplot[color=orange, mark=*, mark size=0.6pt]
    table[x={distance}, y={iter160_flip}] from \DataCSV;
    \addplot[color=red, mark=*, mark size=0.6pt]
    table[x={distance}, y={iter640_flip}] from \DataCSV;

    
    \nextgroupplot[
        align = center,
        ylabel = Different\\Tolerances,
        xmin = -1.0,
        xmax = 20.0,
        xtick distance = 4,
        ymin = 0,
        ymax = 100,
        ytick distance = 25,
        ymajorgrids = true,
        grid style=loosely dashed,
        legend style={
            draw=none,
            anchor=north,   
            at={(1.8,-0.18)},
            legend columns=5,
            /tikz/every even column/.append style={column sep=1.0em},
        },
    ]
    
    \addplot[color=blue, mark=*, mark size=0.6pt]
    table[x={distance}, y={tau512_percentage}] from \DataCSV;
    \addlegendentry{512}
    \addplot[color=cyan, mark=*, mark size=0.6pt]
    table[x={distance}, y={tau1024_percentage}] from \DataCSV;
    \addlegendentry{1024}
    \addplot[color=green, mark=*, mark size=0.6pt]
    table[x={distance}, y={tau2048_percentage}] from \DataCSV;
    \addlegendentry{\underline{2048}}
    \addplot[color=orange, mark=*, mark size=0.6pt]
    table[x={distance}, y={tau4096_percentage}] from \DataCSV;
    \addlegendentry{4096}
    \addplot[color=red, mark=*, mark size=0.6pt]
    table[x={distance}, y={tau8192_percentage}] from \DataCSV;
    \addlegendentry{8192}


    \nextgroupplot[
        xmin = -1.0,
        xmax = 20.0,
        xtick distance = 4,
        ymin = 0,
        ymajorgrids=true,
        grid style=loosely dashed,
    ]
    
    \addplot[color=blue, mark=*, mark size=0.6pt]
    table[x={distance}, y={tau512_rate}] from \DataCSV;
    \addplot[color=cyan, mark=*, mark size=0.6pt]
    table[x={distance}, y={tau1024_rate}] from \DataCSV;
    \addplot[color=green, mark=*, mark size=0.6pt]
    table[x={distance}, y={tau2048_rate}] from \DataCSV;
    \addplot[color=orange, mark=*, mark size=0.6pt]
    table[x={distance}, y={tau4096_rate}] from \DataCSV;
    \addplot[color=red, mark=*, mark size=0.6pt]
    table[x={distance}, y={tau8192_rate}] from \DataCSV;

    
    \nextgroupplot[
        xmin = -1.0,
        xmax = 20.0,
        xtick distance = 4,
        ymin = 0,
        ymax = 0.120,
        ytick distance = 0.040,
        scaled ticks=false,
        y tick label style={/pgf/number format/fixed, /pgf/number format/precision=4, /pgf/number format/fixed zerofill},
        ymajorgrids=true,
        grid style=loosely dashed,
    ]
    
    \addplot[color=blue, mark=*, mark size=0.6pt]
    table[x={distance}, y={tau512_flip}] from \DataCSV;
    \addplot[color=cyan, mark=*, mark size=0.6pt]
    table[x={distance}, y={tau1024_flip}] from \DataCSV;
    \addplot[color=green, mark=*, mark size=0.6pt]
    table[x={distance}, y={tau2048_flip}] from \DataCSV;
    \addplot[color=orange, mark=*, mark size=0.6pt]
    table[x={distance}, y={tau4096_flip}] from \DataCSV;
    \addplot[color=red, mark=*, mark size=0.6pt]
    table[x={distance}, y={tau8192_flip}] from \DataCSV;

    \end{groupplot}
\end{tikzpicture}

\vspace{-10pt}

\caption{\label{fig:ablations-mine-forest-graph}
    \textbf{Graphs from ablation studies.}
    We report metrics comparing renders of 3DGS and our system under varying configurations at twenty relative distances on the composed scene \textsc{Forest}.
}
\Description{ablations-mine-trees}
\end{figure*}
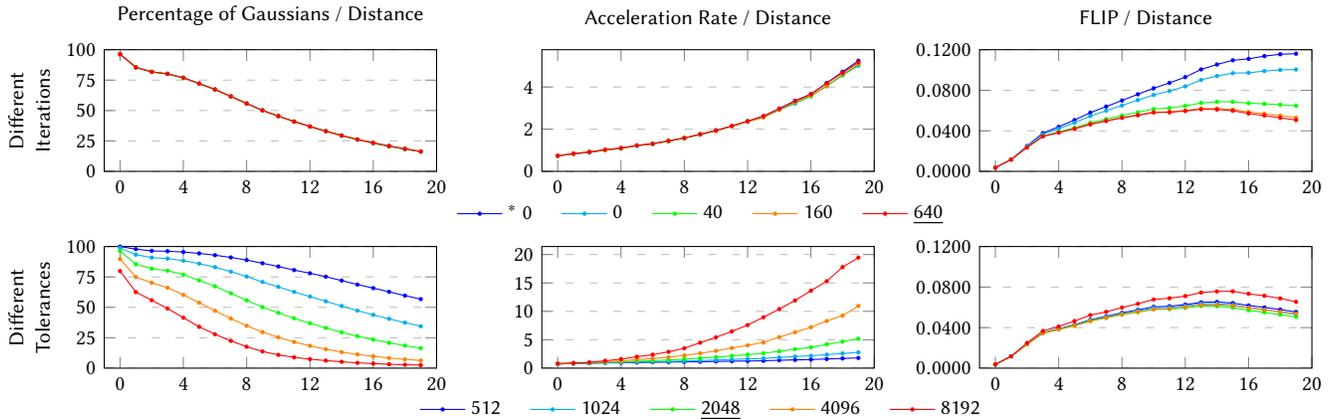


To investigate the influence of different modules within our LOD system, we designed and conducted a series of ablation experiments on \textsc{Forest}.
Table~\ref{tab:ablations-mine-forest} presents the average metrics on all 400 renders, and Fig.~\ref{fig:ablations-mine-forest-graph} displays quantitative metrics at various relative distances.
Additionally, Fig.~\ref{fig:ablation-visual} provides visual comparisons for a specific camera view.
In the table and figures, we underline the basic setting: simplification iteration \underline{640} and selection tolerance \underline{2048}.

\subsubsection{Effectiveness of Local Splatting\label{sec:ablations-local-splatting}}

Local splatting applied within each cluster group is essential for optimizing the properties of simplified 3D Gaussians, ensuring they remain as similar as possible to the original 3D Gaussians.
We evaluate the following five settings to validate the effectiveness of local splatting.

As the number of simplification iteration in the local splatting process increases (0, 40, 160, 640), the percentages of 3D Gaussians and acceleration rates remain largely unchanged, while the FLIP error decreases significantly (Table~\ref{tab:ablations-mine-forest}).
In the upper half of Fig.~\ref{fig:ablation-visual}, the appearance of objects (highlighted in yellow and red frames) rendered using our system becomes closer to that from 3DGS-SSAA with increased iteration, indicating that more iteration of the local splatting leads to improved global rendering quality.
Besides, although larger simplification iteration maintain better fidelity during online rendering, it results in an increased offline build duration, as shown in Table~\ref{tab:ablations-mine-forest}.

We also compared iteration settings $^{\ast}$0 and 0 to explore the influence of initialization during simplification, with scales of initial simplified 3D Gaussians unchanged or increased to $2^{1/6}$ times.
As illustrated in Fig.~\ref{fig:ablation-visual}, instances rendered in the $^{\ast}$0 setting appear sparser after the offline build stage, whereas the increased scales in the 0 setting compensate for this discrepancy.
Additionally, as shown in Fig.~\ref{fig:ablations-mine-forest-graph}, increase in scale leads to a reduction for the FLIP error, indicating that this method provides a more effective initialization.

\begin{figure*}[t]
    \centering
    \includegraphics[width=\linewidth]{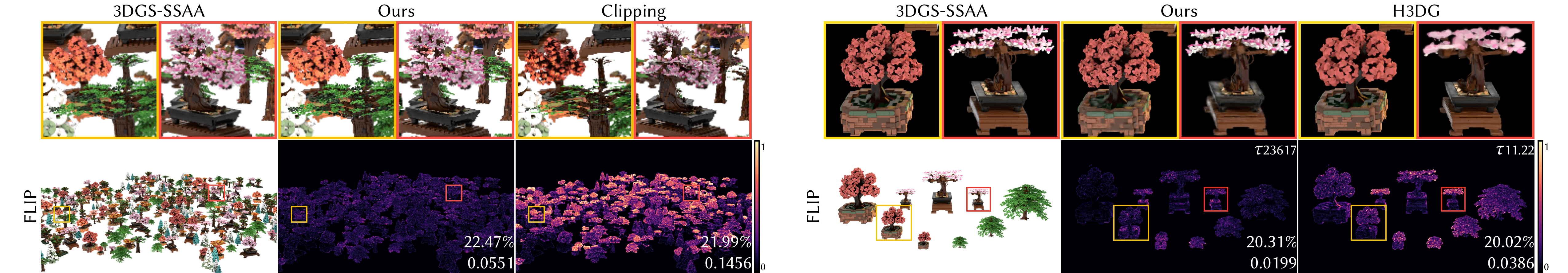}
    \caption{\label{fig:comparison-visual}
        \textbf{Qualitative comparisons with baselines.}
        We show the renders and FLIP \textbf{under the same percentages of selected 3D Gaussians}.
        \emph{Left}: Comparisons between our system and \textbf{gsplat's clipping strategy} on \textsc{Forest}.
        \emph{Right}: Comparisons between our system and \textbf{H3DG} on \textsc{SimpleForest}.
    }
    \Description{comparison-visual}
\end{figure*}

\subsubsection{Different Footprint Tolerances for Selection\label{sec:ablations-footprint-tolerances}}

Footprint tolerance $\tau$ serves as a threshold for determining which clusters to render.
We present different results derived from various values of $\tau$ (512, 1024, 2048, 4096, and 8192) for our ablation study.

As illustrated in Table~\ref{tab:ablations-mine-forest}, Fig.~\ref{fig:ablation-visual} and Fig.~\ref{fig:ablations-mine-forest-graph}, a smaller $\tau$ tends to select finer clusters, until selecting all the finest clusters, thereby maintaining better visual fidelity. Conversely, a larger $\tau$ selects coarser clusters, resulting in a reduced percentage of selected 3D Gaussians and a significantly improved acceleration rate.

Selecting a moderate $\tau$ is crucial for balancing visual quality and rendering speed.
For the composed scene \textsc{Forest}, we choose $\tau=2048$ for our experiments.
However, this value can be flexibly and dynamically adjusted during the online rendering process, allowing digital artists and users to choose appropriate $\tau$ values for rendering the digital world based on their specific requirements for high visual quality or real-time rendering performance.


\subsection{Comparisons\label{sec:comparisons}}

\subsubsection{Comparison with Clipping Strategy in gsplat\label{sec:comparison-radius-clip-in-gsplat}}

To accelerate the rendering of large scenes, gsplat~\cite{gsplat} introduces the argument \textit{radius\_clip}~\footnote{\url{https://docs.gsplat.studio/main/apis/rasterization.html}} to exclude 3D Gaussians with screen-space radius smaller than the given value, thus reducing the total number of 3D Gaussians to rasterize. Using the composed scene \textsc{Forest}, we compare two renders from our system with that produced using gsplat's clipping strategy, as shown in Fig.~\ref{fig:comparison-visual}.
The clipping value is set to 2.0 to maintain comparable percentages ($\sim\!22\%$).

Noticeable cavities appear in the image rendered with the clipping strategy which only chooses \textit{parts of objects} to rasterize. In contrast, our system consistently includes \textit{all parts of full objects at varying levels of detail}, with some newly created through local splatting during the build stage.
Additionally, as the clipping value increases or as the scene is zoomed out, objects can disappear entirely using the clipping strategy. Our system, however, ensures at least one cluster group remains selected to prevent object disappearance.

\subsubsection{Comparison with H3DG\label{sec:comparison-hierarchical-Gaussians}}

Hierarchical 3D Gaussians (H3DG) \cite{hierarchical-gs} introduces a hierarchical 3D Gaussian representation as the core for reconstructing and rendering very large scenes efficiently, sharing similarities with our method.
We carry out two experiments to compare with them: (1) applying their method on our composed scene \textsc{SimpleForest}; (2) applying our method on their standalone chunks of large scenes \textsc{SmallCity} and \textsc{Campus}.

Qualitative results of the first experiment are shown in Fig.~\ref{fig:comparison-visual}.
We compare the two methods using a similar number of 3D Gaussians ($\sim\!20\%$) to render. Our method preserves asset appearance effectively, while H3DG introduces inflated 3D Gaussians, leading to higher FLIP error.
Note that we remove alpha mask multiplication in the training script for H3DG to prevent the appearance of additional protruding artifacts.

\begin{table}[t]
    \caption{\label{tab:comparison-hg}
        \textbf{Metrics comparing with H3DG} on the composed scene \textsc{SimpleForest}.
        Average metrics for renders of the two methods under 120 tolerances with similar numbers of selected 3D Gaussians are presented.
    }

    
    \begin{tabular}{ r | rl }
    \toprule
    Method                    & H3DG   & Ours   \\
    \hline
    Build Duration    $\downarrow$ & 13.9m & 10.5m \\
    (rasterization) FPS $\uparrow$ & 153.69 & 292.83 (1.90x) \\
    (full) FPS         $\uparrow$  & 33.23  & 190.04 (5.71x) \\
    FLIP              $\downarrow$ & 0.0315 & 0.0170         \\
    \bottomrule
    \end{tabular}
\end{table}

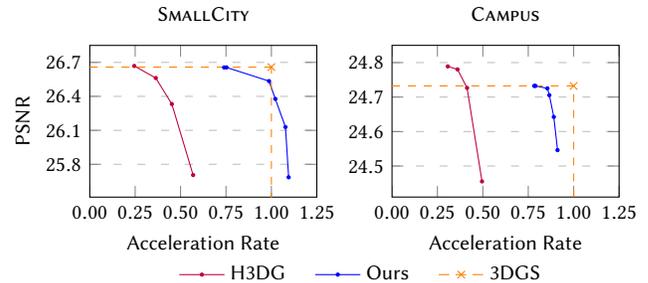
\begin{figure}[t]

\pgfplotstableread[row sep=\\]{
    acceleration_rate_gs psnr_gs \\
    1.0 26.658419640858966 \\
}\DataGSSmallCity

\pgfplotstableread[row sep=\\]{
    fps_gs fps_hiergs psnr_hiergs \\
    123.35783120757677 30.237974952814202 26.668544578552247 \\
    122.69542067351158 44.603857661946044 26.561911710103352 \\
    122.74265990073935 55.460147260487666 26.332425944010417 \\
    123.88158645427826 70.50430686660437 25.704919115702315 \\
}\DataHGSmallCity

\pgfplotstableread[row sep=\\]{
    fps_gs fps_vg psnr_vg \\
    112.7801829830703 83.43371704928975 26.655492464701336 \\
    112.36707771813202 84.56983697994318 26.655491828918457 \\
    108.3052657096242 106.89571984830931 26.534943262736004 \\
    109.13125219196048 111.48772638952686 26.377533785502116 \\
    109.80285695103281 118.30698602752923 26.12860527038574 \\
    110.65513995630207 121.16233023824583 25.685313987731934 \\
}\DataVGSmallCity

\pgfplotstableread[row sep=\\]{
    acceleration_rate_gs psnr_gs \\
    1.0 24.732269446055092 \\
}\DataGSCampus

\pgfplotstableread[row sep=\\]{
    fps_gs fps_hiergs psnr_hiergs \\
    94.05394912892575 28.860032324991057 24.788944721221924 \\
    92.99524042796384 33.53761110887117 24.780205647150677 \\
    93.21885699167622 38.51430522973588 24.72650448481242 \\
    91.11397597446172 45.14096459134481 24.455289045969646 \\
}\DataHGCampus

\pgfplotstableread[row sep=\\]{
    fps_gs fps_vg psnr_vg \\
    85.55981489144575 67.01424771899723 24.7323424021403 \\
    86.94362567190649 68.68980659812676 24.7323424021403 \\
    86.31151004389494 73.82449347814696 24.725210030873615 \\
    84.61773997990475 73.29487789172397 24.70534284909566 \\
    84.22325980780802 75.01055108274782 24.642447471618652 \\
    82.61447632471518 75.27971492312237 24.54613709449768 \\
}\DataVGCampus

\begin{tikzpicture}[font=\small\sffamily]
    \begin{groupplot}[
        group style = {
            group size = 2 by 1,
            horizontal sep = 1cm,
        },
        height=3.6cm,
        width=4.6cm,
    ]
    
    \nextgroupplot[
        title = \textsc{SmallCity},
        xlabel = Acceleration Rate,
        ylabel = PSNR,
        xmin = 0.0,
        xmax = 1.25,
        xtick distance = 0.25,
        x tick label style={/pgf/number format/fixed, /pgf/number format/precision=2, /pgf/number format/fixed zerofill},
        ymin = 25.51,
        ymax = 26.85,
        ytick distance = 0.3,
        y tick label style={/pgf/number format/fixed, /pgf/number format/precision=1, /pgf/number format/fixed zerofill},
        ymajorgrids = true,
        grid style=loosely dashed,
        legend style={
            draw=none,
            anchor=north,   
            at={(1.2,-0.38)},
            legend columns=-1,
            /tikz/every even column/.append style={column sep=1.0em},
        },
    ]
    \addplot[color=purple, mark=*, mark size=0.6pt]
    table[x expr=\thisrow{fps_hiergs} / \thisrow{fps_gs}, y={psnr_hiergs}] from \DataHGSmallCity;
    \addlegendentry{H3DG}
    \addplot[color=blue, mark=*, mark size=0.6pt]
    table[x expr=\thisrow{fps_vg} / \thisrow{fps_gs}, y={psnr_vg}] from \DataVGSmallCity;
    \addlegendentry{Ours}
    \addplot[color=orange, mark=x, mark size=2pt, dashed, dash pattern=on 4pt off 2.4pt]
    table[x={acceleration_rate_gs}, y={psnr_gs}] from \DataGSSmallCity;
    \addlegendentry{3DGS}
    \addplot[dashed, orange] coordinates {(1.0, 0.0) (1.0, 26.658419640858966)};
    \addplot[dashed, orange] coordinates {(0.0, 26.658419640858966) (1.0, 26.658419640858966)};


    \nextgroupplot[
        title = \textsc{Campus},
        xlabel = Acceleration Rate,
        xmin = 0.0,
        xmax = 1.25,
        xtick distance = 0.25,
        x tick label style={/pgf/number format/fixed, /pgf/number format/precision=2, /pgf/number format/fixed zerofill},
        ymin = 24.41,
        ymax = 24.85,
        ytick distance = 0.1,
        y tick label style={/pgf/number format/fixed, /pgf/number format/precision=1, /pgf/number format/fixed zerofill},
        ymajorgrids = true,
        grid style=loosely dashed,
    ]
    \addplot[color=purple, mark=*, mark size=0.6pt]
    table[x expr=\thisrow{fps_hiergs} / \thisrow{fps_gs}, y={psnr_hiergs}] from \DataHGCampus;
    \addplot[color=blue, mark=*, mark size=0.6pt]
    table[x expr=\thisrow{fps_vg} / \thisrow{fps_gs}, y={psnr_vg}] from \DataVGCampus;
    \addplot[color=orange, mark=x, mark size=2pt, dashed, dash pattern=on 4pt off 2.4pt]
    table[x={acceleration_rate_gs}, y={psnr_gs}] from \DataGSCampus;
    \addplot[dashed, orange] coordinates {(1.0, 0.0) (1.0, 24.732269446055092)};
    \addplot[dashed, orange] coordinates {(0.0, 24.732269446055092) (1.0, 24.732269446055092)};
    
    \end{groupplot}
\end{tikzpicture}

\vspace{-10pt}

\caption{\label{fig:comparison-hiergs-graph}
    \textbf{Comparison with H3DG on two standalone chunks} of \textsc{SmallCity} and \textsc{Campus} provided by \cite{hierarchical-gs}.
    In each graph, from left to right: H3DG ($\tau = 0, 3, 6, 15$) and Ours ($\tau = 0, 2^8, 2^{16}, 2^{17}, 2^{18}, 2^{19}$).
}
\Description{comparison-hiergs-graph}

\end{figure}

We also provide a supplementary video showcasing continuously changing tolerances. While H3DG interpolation ensures smooth transitions, inflated artifacts diminish its advantages. The video includes renders of unoptimized H3DG hierarchies and detailed FPS statistics. Average metrics in Table~\ref{tab:comparison-hg} demonstrate that our method delivers superior visual quality and greater efficiency.

The quantitative results of the second experiment are presented in Fig.~\ref{fig:comparison-hiergs-graph}. PSNRs of test views are computed, as standalone chunks in H3DG include captured ground truths. Under the same rendering quality, our system achieves comparable speed to 3DGS and is approximately twice as fast as H3DG. This is primarily because our cluster selection stage introduces minimal overhead before rasterization, whereas H3DG's interpolation process significantly increases the overhead of both selection and rasterization.

\section{CONCLUSION AND DISCUSSION\label{sec:conclusion-and-discussion}}

In this work, we introduce Virtualized 3D Gaussians (V3DG), a cluster-level LOD system designed for real-time rendering of a digital world containing numerous 3DGS assets.
The core of our system lies in the novel offline build and online selection stages for 3D Gaussians, which are inspired by Nanite, selectively rendering only perceivable 3D Gaussians and no more through a cluster-based design.
Our experiments demonstrate that the proposed LOD system effectively employs a minimal number of 3D Gaussian primitives and accelerates the rendering process while maintaining the visual fidelity of 3DGS assets in the digital world.

However, one limitation of our method is its increased storage requirement, nearly doubling the space needed to store pre-built bundles for accelerated online rendering. This overhead could potentially be mitigated by integrating existing \textit{compression techniques} for 3DGS.
Regarding memory usage, although our system is described as "Virtualized" 3D Gaussians, it currently performs adaptive 3D Gaussian selection in memory but lacks persistent disk-to-memory streaming. To enable more efficient memory management, a streaming module is needed to dynamically load selected clusters into memory while periodically discarding unused ones.
Additionally, the composed scenes used in our evaluation remain relatively simple. Future evaluations on production-scale digital environments would provide a more rigorous assessment of our system’s capabilities.
We leave addressing these limitations and exploring possible improvements to future work.


\begin{acks}

This work was partially supported by the National Key R\&D Program of China (2022ZD0160201), the Shanghai Artificial Intelligence Laboratory, and the HKU Startup Fund. We would also like to thank Kerui Ren and Mulin Yu for their insightful discussions, and the anonymous reviewers for their valuable feedback.

\end{acks}


\bibliographystyle{ACM-Reference-Format}
\bibliography{parts/references}



\clearpage
\appendix

The following appendices provide additional technical details and experimental results that support the main findings of this work.
They include descriptions of the dataset preparation and asset acquisition process (\ref{sec:appendices-datasets-acquisition-of-3dgs-assets}), the construction of composed scenes (\ref{sec:appendices-datasets-construction-of-composed-scenes}), and dynamic scene (\ref{sec:appendices-datasets-dynamic-scene}).
Detailed analyses on the system’s performance, including visualizations (\ref{sec:appendices-additional-analysis-visualization-across-layers}), render quality (\ref{sec:appendix-detailed-analysis-on-renders}), and comparisons with existing methods like H3DG (\ref{sec:appendices-additional-analysis-theoretical-caomparisons-with-h3dg}), are also provided.
Additionally, supplementary experiments explore the effects of hierarchical structures (\ref{sec:appendices-ablation-dag-and-tree}), large scene rendering (\ref{sec:appendices-large-scenes}), and results on original views of the NVS datasets (\ref{sec:appendices-experiments-single-assets}).
These appendices offer a deeper understanding of the methodology and provide transparency into the experimental setup and results.


\begin{algorithm}[b]
    \caption{\textbf{Offline Build Stage}\label{alg:lod-build-stage}\\
    \textbf{input}: $N_G$ 3D Gaussians $G$ in a 3DGS asset $A$.\\
    \textbf{output}: a bundle $B$ with clusters $C_0, C_1, \dots, C_m$ in all layers, where $m = \lceil \log_2{(N_G / (n_{G \in C} \cdot n_{C \in C}))} \rceil$.\\
    $A$: a given 3DGS \underline{a}sset.\\
    $G$: 3D \underline{G}aussians in $A$.\\
    $i$: \underline{i}d of current layer.\\
    $C_i$: all \underline{c}lusters in layer \underline{$i$}.\\
    $CG_i$: all \underline{c}luster \underline{g}roups in layer \underline{$i$}.\\
    $C_o, G_o$: \underline{o}riginal \underline{c}lusters and 3D \underline{G}aussians in a cluster group.\\
    $G_s, C_s$: \underline{s}implified 3D \underline{G}aussians and \underline{c}lusters in a cluster group.\\
    $N_G$: count of 3D \underline{G}aussians.\\
    $N_C$: count of \underline{c}lusters.\\
    $n_{G \in C}$: count of 3D \underline{G}aussians in a \underline{c}luster.\\
    $n_{C \in CG}$: count of \underline{c}lusters in a \underline{c}luster \underline{g}roup.\\
    $B$: the output bundle with discrete clusters.}
    \begin{algorithmic}
        \State $G\ \textbf{in}\ A$ \Comment{\AlgSymbol{$N_G$}}
        \State $C_0 \gets$ Split($G$, $N_G / n_{G \in C}$) \Comment{\AlgSymbol{$N_{C_0}$}} \\ \Comment{\AlgComment{split 3D Gaussians into clusters}}
        \State $B \overset{+}{\gets} C_0$ \Comment{\AlgComment{initialize bundle}}
        \State $i \gets 0$
        \While{$N_{C_i} \geq n_{C \in CG}$} \Comment{\AlgComment{build coarser layer}}
            \State $CG_i \gets$ Split($C_i$, $N_{C_i} / n_{C \in CG}$) \\ \Comment{\AlgComment{split clusters into cluster groups}}
            \ForAll{$CG\ \textbf{in}\ CG_i$} \Comment{\AlgComment{simplify in each cluster group}}
                \State $C_o\ \textbf{in}\ CG$ \Comment{\AlgSymbol{$N_{C_o}$}}
                \State $G_o\ \textbf{in}\ C_o$ \Comment{\AlgSymbol{$N_{G_o}$}}
                \State $G_s \gets$ Simplify($G_o$) \Comment{\AlgSymbol{$N_{G_s} \approx \frac{1}{2} N_{G_o}$}} \\ \Comment{\AlgComment{local splatting method shown in Sec.~\ref{sec:method-simplification}}}
                \State $C_s \gets$ Split($G_s$, $N_{G_s} / n_{G \in C}$) \Comment{\AlgSymbol{$N_{C_s} \approx \frac{1}{2} N_{C_o}$}} \\ \Comment{\AlgComment{split simplified 3D Gaussians into clusters}}
                \State $C_{i+1} \overset{+}{\gets} C_s$
            \EndFor
            \State $B \overset{+}{\gets} C_{i+1}$ \Comment{\AlgSymbol{$N_{C_{i+1}} \approx \frac{1}{2} N_{C_i}$}} \\ \Comment{\AlgComment{consolidate clusters into bundle}}
            \State $i \gets i+1$
        \EndWhile
    \end{algorithmic}
\end{algorithm}


\section{DATASET PREPARATION\label{sec:appendices-datasets}}


\subsection{Acquisition of 3DGS Assets\label{sec:appendices-datasets-acquisition-of-3dgs-assets}}

To derive the 3DGS assets used in our dataset, we collect multiple NVS datasets and apply a 3D reconstruction algorithm.

\subsubsection{NVS Datasets}

For object-level assets, we prepare \textsc{donut}, \textsc{trees}, and \textsc{persons}. For \textsc{donut}, a 3D model downloaded from the Internet is imported into Blender, rendered at 800x800 resolution, and exported following the dataset creation process in NeRF~\cite{nerf}. For \textsc{trees}, 8 representative objects are selected from the RTMV Bricks dataset~\cite{RTMV}, with renders produced at 1600x1600 resolution. For \textsc{persons}, we choose 16 identities from the MVHumanNet dataset~\cite{mvhumannet}, using 48 views from the first frame at a resolution of 2048x1500.

For scene-level assets, we prepare \textsc{matrixcity}, \textsc{downtown}, and \textsc{temple}. \textsc{Matrixcity} uses aerial captures from the MatrixCity dataset \cite{matrixcity} at 1920x1080 resolution. \textsc{Downtown} and \textsc{temple} are captured directly from real environments, with resolutions of 1900x1267 and 1638x1092, respectively.

\subsubsection{Reconstruction Algorithm}

The reconstruction algorithm applied to the NVS datasets incorporates several modifications based on 3DGS~\cite{gs} and Mip-Splatting~\cite{mipgs}. We set the 2D Mip filter to 0.0, ensuring that the scale of each 3D Gaussian accurately represents its 3D occupancy. This adjustment addresses a missing normalization in the original 3DGS implementation, which can cause dilation and erosion artifacts at varying distances~\cite{mipgs}. Additionally, a 3D smoothing filter value of 0.1 is applied to prevent the generation of excessively small 3D Gaussians, stabilizing the optimization process during the offline build stage. After reconstruction, the 3D filter values are fused into the properties of the 3D Gaussians to generate the ply model in vanilla 3DGS.

For memory efficiency and simplicity, only 0-degree spherical harmonics are used to represent the view-independent colors of the 3D Gaussians. Since the object-level NVS datasets contain alpha channel information, we use the groundtruths with alpha channels directly in training, avoiding the need for background blending. The alpha channel supervision ($\mathcal{L} \mathrel{+}= 0.1 \mathcal{L}_{1} (I_{\alpha \ \text{render}}, I_{\alpha \ \text{groundtruth}})$, where $I_\alpha$ is the accumulated opacity image) enhances reconstruction quality by reducing floating artifacts.

For specific datasets, we adjust parameters to optimize performance. In the case of \textsc{donut}, we set the 3D smoothing filter to 0.001 to avoid blurring, given the model's small scale. For \textsc{persons}, we increase the densification and reduce the number of training iterations to prevent overfitting. In \textsc{matrixcity}, a dense and colorful point cloud from depth information is used for initialization, without the adaptive control in vanilla 3DGS. For the scene-level assets \textsc{matrixcity}, \textsc{downtown}, and \textsc{temple}, we extend the training to 120,000 iterations to improve reconstruction quality. Following reconstruction, the three scene-level assets are manually cropped to focus on their main components, enhancing visualization and composition.

\subsubsection{Reconstruction Records}

\begin{table*}[t]
    \caption{\label{tab:appendices-datasets-composed-scenes}
        \textbf{Details of the four composed scenes in our dataset.}
        This table provides the number of assets, instances, and 3D Gaussians in each composed scene used in our experiments, accompanied by construction notes for each scene.
    }
    \begin{tabular}{ r | cccl }
    \toprule
    Composed Scenes   & \#Assets & \#Instances & \#Gaussians & Notes \\
    \hline
    \textsc{DonutSea} & 1       & 2,048      & 93,437,952  & physical simulation, random scale \\
    \textsc{Forest}   & 8       & 400        & 125,917,439 & heuristic layout, random 2D rotation and scale \\
    \textsc{Crowd}    & 16      & 400        & 133,504,977 & grid layout, fixed 2D rotation and scale \\
    \textsc{Temples}  & 1       & 100        & 135,048,700 & heuristic layout, random 2D rotation and scale \\
    \bottomrule
    \end{tabular}
\end{table*}

\begin{table}[t]
    \caption{\label{tab:appendices-datasets-assets-nvs-metrics}
        \textbf{Metrics on views in NVS datasets.}
        We evaluate vanilla 3DGS and our system across all 3DGS assets used in our experiments.
        Statistics for cropped assets are \textit{italicized}.
    }
    \begin{tabular}{ cr | ccccccc }
    
    \toprule
    
    & \multirow{2}{*}{Object-Level Assets} &  & average & average \\
    & & \textsc{donut} & \textsc{trees} & \textsc{persons} \\
    
    \hline
    
    \parbox[t]{2mm}{\multirow{5}{*}{\rotatebox[origin=c]{90}{3DGS}}}
    & Recon. Duration & 9min & 15min & 4min  \\
    & Total Gaussians   & 45,624 & 315,630 & 343,274 \\
    & Visible Gaussians & 45,624 & 267,898 & 343,274 \\
    & FPS               & 665    & 283     & 236     \\
    & PSNR              & 39.97  & 31.70   & 40.48   \\

    \hline
    
    \parbox[t]{2mm}{\multirow{4}{*}{\rotatebox[origin=c]{90}{Ours}}}
    & Build Duration     & 1min     & 10min    & 11min   \\
    & Selected Gaussians & 100.00\% & 100.00\% & 99.05\% \\
    & Acceleration Rate  & 0.45x    & 0.63x    & 0.70x   \\
    & PSNR               & 39.97    & 31.70    & 40.48   \\
    
    \midrule
    
    & \multirow{2}{*}{Scene-Level Assets} & \multicolumn{1}{c}{synthetic} & \multicolumn{2}{c}{real} \\
    & & \textsc{matrixcity} & \textsc{downtown} & \textsc{temple} \\
    
    \hline

    \parbox[t]{2mm}{\multirow{6}{*}{\rotatebox[origin=c]{90}{3DGS}}}
    & Recon. Duration & 854min & 346min & 240min \\
    & Total Gaussians         & 15,540,941 & 15,407,110 & 9,269,300 \\
    & \textit{Total Gaussians} & \textit{12,925,152} & \textit{7,754,285} & \textit{1,350,487} \\
    & Visible Gaussians       & 447,086 & 673,825 & 2,234,318 \\
    & FPS                     & 155 & 85 & 60 \\
    & PSNR                    & 27.39 & 24.50 & 25.50 \\
    
    \hline
    
    \parbox[t]{2mm}{\multirow{5}{*}{\rotatebox[origin=c]{90}{Ours}}}
    & Build Duration     & 378min & 417min & 398min \\
    & \textit{Build Duration} & \textit{367min} & \textit{188min} & \textit{44min} \\
    & Selected Gaussians & 99.98\% & 99.51\% & 82.94\% \\
    & Acceleration Rate  & 0.73x & 0.77x & 0.97x \\
    & PSNR               & 27.39 & 24.50 & 25.42 \\

    \bottomrule
    
    \end{tabular}
\end{table}


Table~\ref{tab:appendices-datasets-assets-nvs-metrics} presents reconstruction metrics for asset acquisition from the NVS datasets, including reconstruction duration, total number of 3D Gaussians in the ply file, average number of visible 3D Gaussians within camera frustums, and PSNR relative to the groundtruths in NVS datasets.


\subsection{Construction of Composed Scenes\label{sec:appendices-datasets-construction-of-composed-scenes}}

We construct four composed scenes, \textsc{DonutSea}, \textsc{Forest}, \textsc{Crowd}, and \textsc{Temples}, from the derived assets.

For the scene \textsc{DonutSea}, we simulate the physical process of donuts falling onto a plane with random scales, rotations, and positions using Blender, then export the fallen donuts to form the scene.
The \textsc{Forest} and \textsc{Temples} scenes are generated using a procedural script that incrementally places asset instances on the plane with random scales and 2D rotations. The \textsc{Crowd} scene is created by manually arranging multiple instances of \textsc{persons} in a grid layout on a 2D plane.

Table~\ref{tab:appendices-datasets-composed-scenes} lists the number of instances and 3D Gaussians for each scene. The total number of 3D Gaussians is approximately 0.1 billion per scene, as GPU memory constraints limit loading scenes with billions of 3D Gaussians. While our system is designed to handle larger scales, a streaming module, as discussed in Sec.~\ref{sec:conclusion-and-discussion}, is required to render scenes with such a high number of primitives.


\subsection{Dynamic Scenes\label{sec:appendices-datasets-dynamic-scene}}

\begin{figure}[t]
    \centering
    \includegraphics[width=\linewidth]{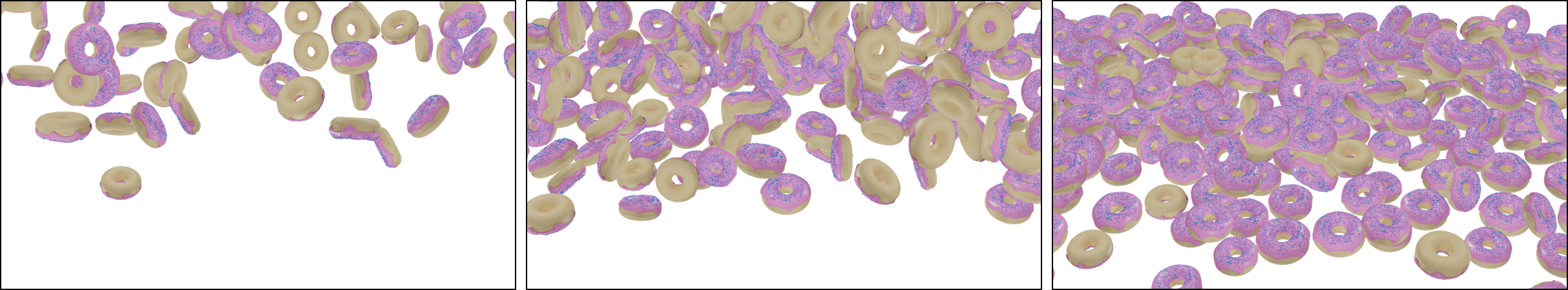}
    \caption{\label{fig:dataset-falling-donuts}
        \textbf{Example of a dynamic scene.}
        This illustration depicts three consecutive frames of donuts falling from the sky. Our system enables the rendering of such dynamic scenes by selecting clusters and rasterizing 3D Gaussians for each individual frame.
    }
    \Description{dataset-falling-donuts}
\end{figure}

Although all the composed scenes in our experiments are static, our system also supports accelerating dynamic scenes with rigid assets. At each frame, our system selects clusters within scene instances, enabling the acceleration of dynamic scenes with moving and rotating rigid objects. For example, Fig.~\ref{fig:dataset-falling-donuts} illustrates a dynamic scene with multiple frames showing donuts falling from the sky.
However, currently, fully dynamic assets (e.g., a robot waving its arms, with the robot treated as a whole asset) are not supported.


\section{Additional Analysis}


\subsection{Visualization across Layers\label{sec:appendices-additional-analysis-visualization-across-layers}}

\begin{figure*}[t]
    \centering
    \includegraphics[width=\linewidth]{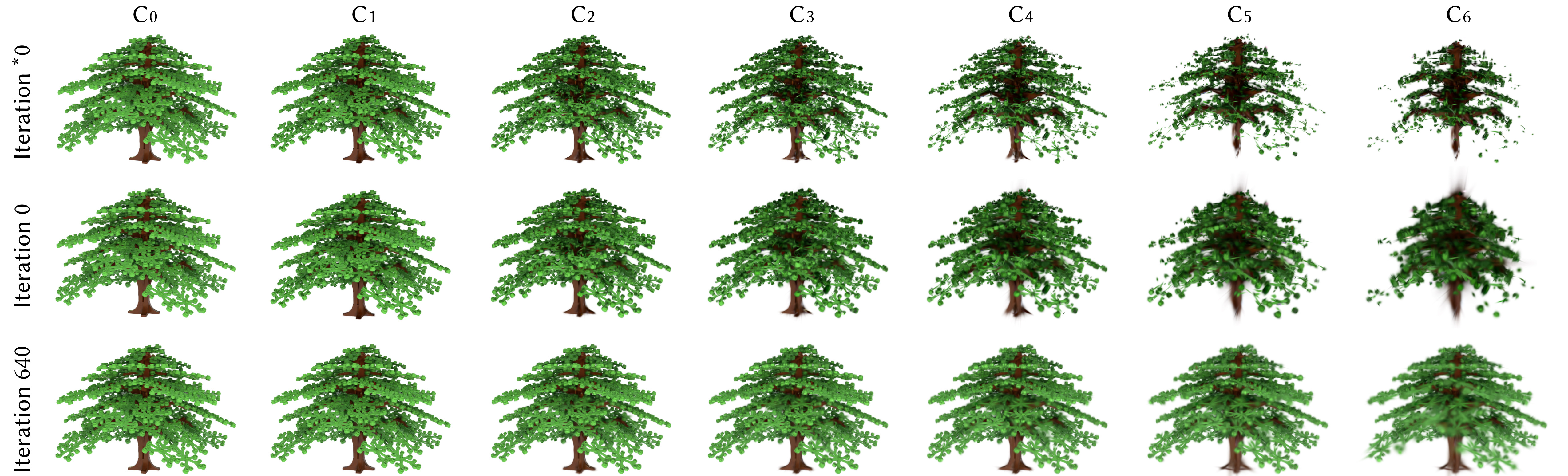}
    \caption{\label{fig:appendix-oak-layers}
        \textbf{Visualization of clusters across layers in a bundle.}
        \textit{From left to right}: Renders of 3D Gaussians in each layer.
        \textit{From top to down}: Different simplification iteration settings.
        This demonstrates that \textbf{scale expansion} during initialization and the \textbf{local splatting} method in simplification are essential for preserving the appearance of 3DGS assets. Notably, our method selects clusters across multiple layers simultaneously, with single-layer cluster selection being rare.
    }
    \Description{appendix-oak-layers}
\end{figure*}

We consolidate 3D Gaussians in clusters by layer and rasterize them for visualization in Fig.~\ref{fig:appendix-oak-layers}. The initialization scale expansion prevents overly thin 3D Gaussians in coarser layers, while the local splatting method maintains the global appearance of the 3D Gaussian asset across all layers.


\subsection{Detailed Analysis on Renders\label{sec:appendix-detailed-analysis-on-renders}}

\begin{figure}[t]
    \centering
    \includegraphics[width=\linewidth]{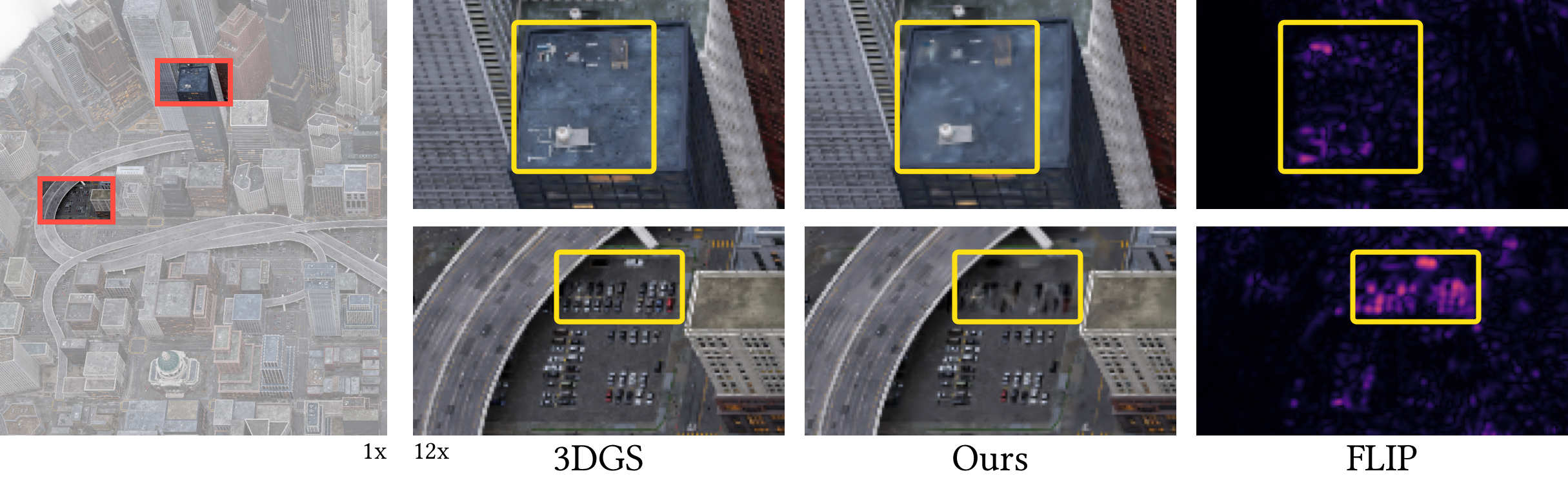}
    \caption{\label{fig:Experiment-high-frequency-suppression}
        \textbf{Possible spatial artifacts produced by our system.}
        Our system may generate artifacts that reduce certain high-frequency details.
    }
    \Description{Experiment-high-frequency-suppression}
\end{figure}

\begin{figure}[t]
    \centering
    \includegraphics[width=\linewidth]{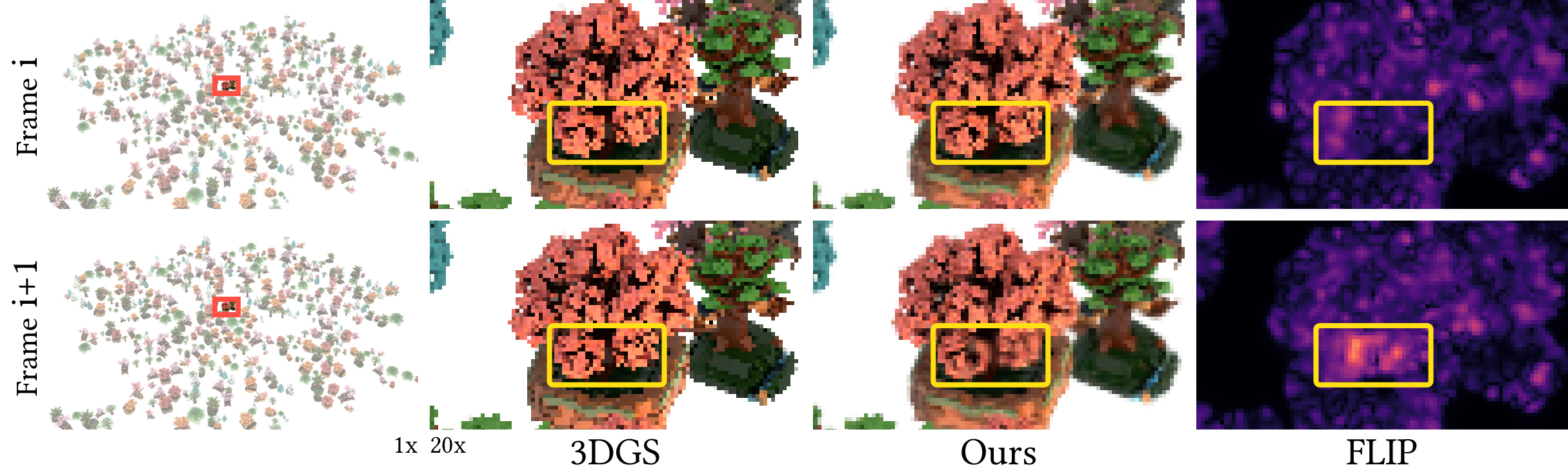}
    \caption{\label{fig:experiment-popping}
        \textbf{Possible temporal artifacts produced by our system.}
        Two consecutive frames from the \textsc{Forest} scene. Alternating between these frames reveals popping artifacts highlighted within the yellow rectangles.
    }
    \Description{experiment-popping}
\end{figure}

We have mentioned the advantages of our system in the body part, including satisfying acceleration rates toward real-time, and the anti-aliasing function for more comfortable visual quality.

While our system achieves real-time acceleration and improved visual quality through anti-aliasing, some artifacts remain in the renders. Fig.~\ref{fig:Experiment-high-frequency-suppression} illustrates the suppression of high-frequency details due to simplification (magnified 12x). Although generally acceptable for entertainment applications, these artifacts may be problematic for applications requiring high precision.

Additionally, the cluster selection process before rasterization introduces a slight FPS drop at close distances, as seen in Fig.\ref{fig:experiment-composed-scenes-graph} and Table~\ref{tab:appendices-datasets-assets-nvs-metrics}. This overhead could potentially be reduced by optimizing the Python code used for selection.

In the supplementary video, aliasing is more pronounced in consecutive 3DGS frames, but our system significantly reduces aliasing, particularly in the \textsc{Temples} scene. However, transitions between adjacent layers may cause slight popping artifacts, as shown in Fig.~\ref{fig:experiment-popping} and the supplementary video. Possible solutions include the interpolation strategy in H3DG~\cite{hierarchical-gs} or Temporal Anti-Aliasing (TAA) used in Unreal Engine's Nanite~\cite{nanite}, which we plan to explore in future work.


\subsection{Theoretical Comparisons with H3DG\label{sec:appendices-additional-analysis-theoretical-caomparisons-with-h3dg}}

HierarchicalGaussians (H3DG) \cite{hierarchical-gs} uses a hierarchical 3D Gaussian representation to reconstruct and render large scenes. The process involves three main steps:
(1) Scene images are captured and divided into chunks, each trained with a modified 3DGS;
(2) Hierarchies are created for each chunk and optimized using the captured images;
(3) The chunk hierarchies are consolidated into a single hierarchy, with LOD rendering accelerated via cut selection based on the camera position. 3D Gaussians on the selected cut are interpolated and rasterized to produce the final image.

We apply H3DG to our composed scene acceleration task, aligning with our offline build stage by generating LOD for each 3DGS asset from multi-view groundtruths (step 2 in H3DG). Similarly, during runtime (step 3), H3DG selects cuts for each 3DGS asset and rasterizes the consolidated 3D Gaussians for LOD-aware rendering. We conduct an experiment comparing our system with H3DG, as detailed in Sec.~\ref{sec:comparison-hierarchical-Gaussians}, and analyze the key differences, advantages, and drawbacks.

\paragraph{Basic Unit.}
In our method, the cluster, a group of local 3D Gaussians, outperforms the single 3D Gaussian used in H3DG. The differentiable nature of 3D Gaussians allows our method to simplify thousands of 3D Gaussians efficiently through local splatting, whereas H3DG requires careful design to consolidate two 3D Gaussians into one when building hierarchies.

\paragraph{Steady Optimization.}
Our method uses a rule-based, uniform approach, optimizing a fixed number of local 3D Gaussians in clusters at a constant resolution (e.g., 64x64). This is scalable and parallelizable. In contrast, H3DG optimizes varying numbers of 3D Gaussians globally at full resolution in each iteration, leading to unstable training performance and serial iterations.

\begin{figure*}[t]
    \centering
    \includegraphics[width=\linewidth]{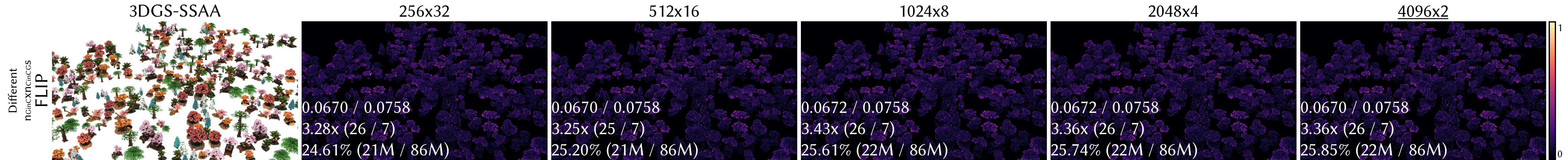}
    \caption{\label{fig:Exploration-dag-and-tree-visual}
        \textbf{Renders from ablation study for varying $n_{G \in C} \times n_{C \in CG}$ configurations} on the composed scene \textsc{Forest}.
        No prominent difference can be observed across these renders.
    }
    \Description{Exploration-dag-and-tree-visual}
\end{figure*}

\paragraph{Post-processing instead of data-driven.}
Unlike H3DG, which relies on data-driven optimization, our method performs general post-processing on 3DGS assets without the need for groundtruths. This is particularly useful for assets like the \textsc{temple} in our dataset, which lacks groundtruths after cropping from a larger scene. Our method is also applicable to 3DGS assets generated by models, whereas H3DG would require additional groundtruths or masks for optimal quality.

\paragraph{Interpolation}
H3DG's interpolation between 3D Gaussians provides smooth LOD transitions, though it comes at the cost of rendering speed (about half of our method at the same image quality). It also requires a well-optimized hierarchy for optimal visual quality. Our method lacks this interpolation mechanism, leading to potential flickering between layers, though this can be mitigated by narrowing the tolerance when switching granularities. Temporal Anti-Aliasing (TAA), used in Nanite~\cite{nanite}, addresses similar issues by blending sub-pixel differences over time, and could be a valuable avenue for future work.



\begin{table}[t]
    \caption{\label{tab:exploration-dag-and-tree}
        \textbf{Metrics from ablation study for varying $n_{G \in C} \times n_{C \in CG}$ configurations} on the composed scene \textsc{Forest}.
        The basic setting is \underline{underlined}.
    }
    \begin{tabular}{ rr | cccc }
    \toprule
    ~ & ~ & Duration & Percentage & Rate & FLIP \\
    \hline
    \parbox[t]{2mm}{\multirow{5}{*}{\rotatebox[origin=c]{90}{$n_{G \in C} \times n_{C \in CG}$}}}
        &  256 $\times$ 32            &10.4m& 33.10\% & 1.47x & 0.0462 \\
        &  512 $\times$ 16            &10.2m& 33.58\% & 1.46x & 0.0462 \\
        & 1024 $\times$ 8             &10.3m& 33.86\% & 1.47x & 0.0462 \\
        & 2048 $\times$ 4             &10.2m& 34.03\% & 1.46x & 0.0462 \\
        & \underline{4096 $\times$ 2} &\underline{9.8m} & \underline{34.28\%} & \underline{1.47x} & \underline{0.0462} \\
    \bottomrule
    \end{tabular}
\end{table}

\section{SUPPLEMENTARY EXPERIMENTS\label{sec:appendices-supplementary-experiments}}


\subsection{Exploration of DAG and Tree\label{sec:appendices-ablation-dag-and-tree}}

In our method, adjacent $n_{G \in C}$ 3D Gaussians form clusters, which act as nodes in the hierarchical LOD system. $n_{C \in CG}$ clusters are further grouped into cluster groups. When $n_{C \in CG} = 1$, the smallest unit in our method will break apart during simplification, whereas larger $n_{C \in CG}$ values define boundaries within which simplification occurs. Smaller values of $n_{C \in CG}$ may increase errors or collects cracks across LOD layers, while larger values risk losing fine details during simplification.
For our ablation study, we keep the total number of 3D Gaussians in each cluster group constant at 8192 by varying $n_{G \in C}$ and $n_{C \in CG}$ values (256 $\times$ 32, 512 $\times$ 16, 1024 $\times$ 8, 2048 $\times$ 4, 4096 $\times$ 2).

In Nanite~\cite{nanite}, a higher $n_{C \in CG}$ (as in a DAG structure) leads to smoother LOD transitions, while the 4096 $\times$ 2 configuration (tree structure) results in denser visual artifacts, or "cruft".
However, in the context of 3D Gaussians, our experiments reveal no significant difference between DAG and tree configurations, as demonstrated in Table~\ref{tab:exploration-dag-and-tree} and Fig.~\ref{fig:Exploration-dag-and-tree-visual}. We attribute this to the inherent flexibility of 3D Gaussian simplification. Unlike mesh-based representations, such as those used in Nanite, where boundaries are rigidly maintained during simplification, our local splatting strategy permits adaptive boundaries. This adaptability helps avoid the accumulation of dense cruft and contributes to more consistent visual quality across levels of detail.


\subsection{Results on Large Scenes\label{sec:appendices-large-scenes}}

\begin{figure*}[t]
    \centering
    \includegraphics[width=\linewidth]{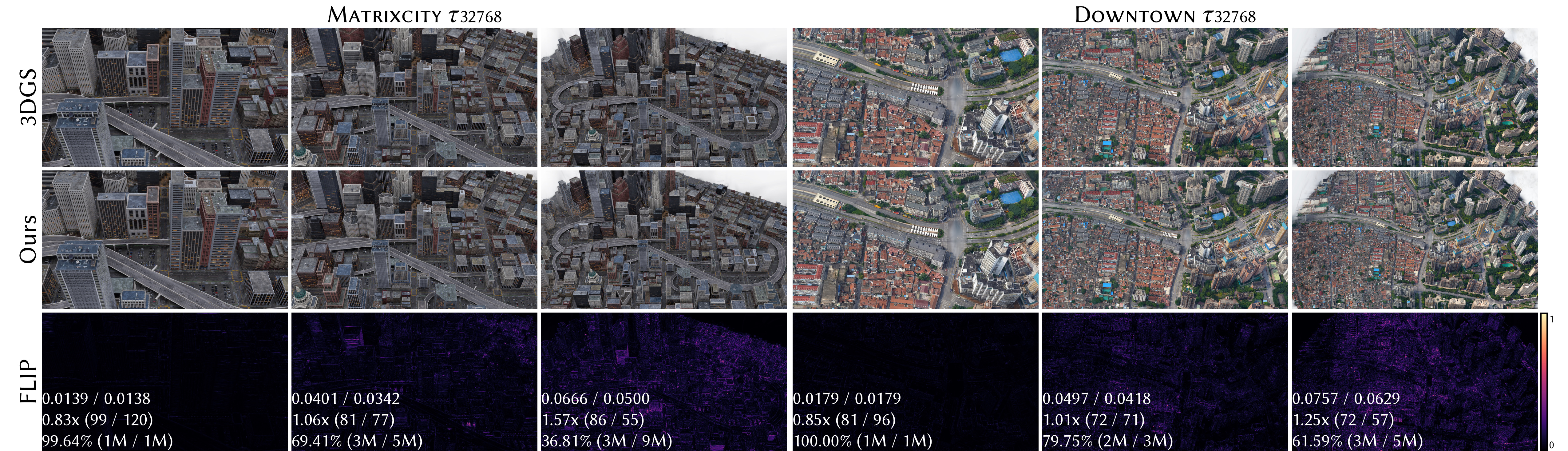}
    \caption{\label{fig:experiment-single-large-visual}
        \textbf{Renders on individual large scenes.}
        We show renders of 3DGS and our system on two large scenes at three different distances.
        FLIP errors, acceleration rates (x), and percentages of selected Gaussians (\%) are indicated at the bottom-left corner (Ours / 3DGS).
    }
    \Description{experiment-single-large-visual}
\end{figure*}

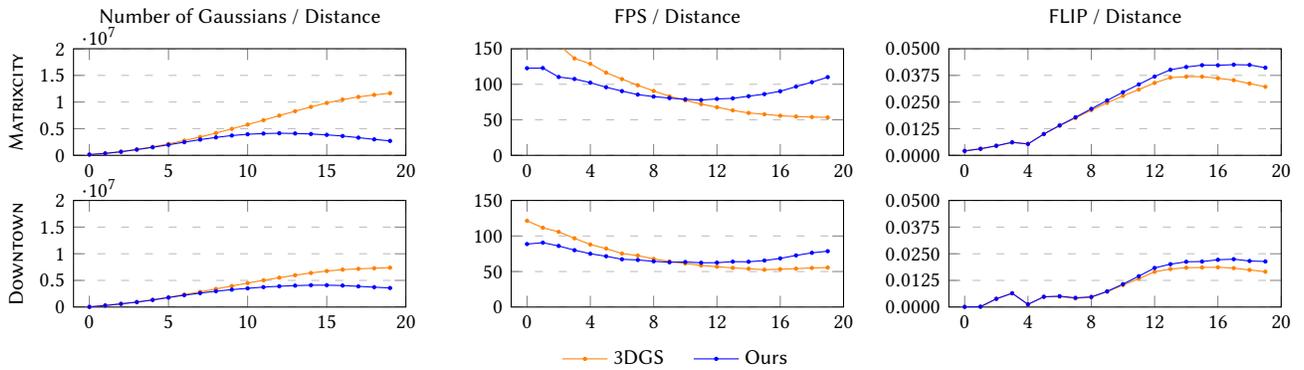
\begin{figure*}[t]

\pgfplotstableread[col sep=comma]{graphs/data/data-Matrixcity.csv}\DataMatrixcity
\pgfplotstableread[col sep=comma]{graphs/data/data-Downtown.csv}\DataDowntown

\begin{tikzpicture}[font=\small\sffamily]
    \begin{groupplot}[
        group style = {
            group size = 3 by 2,
            horizontal sep = 1.4cm,
            vertical sep = 0.6cm,
        },
        height=3.0cm,
        width=6cm
    ]


    \nextgroupplot[
        title = Number of Gaussians / Distance,
        ylabel = \textsc{Matrixcity},
        xmin = -1.0,
        xmax = 20.0,
        ymin = 0,
        ymax = 20000000,
        ymajorgrids = true,
        grid style=loosely dashed,
    ]
    \addplot[color=orange, mark=*, mark size=0.6pt]
    table[x={distance}, y={gs_count}] from \DataMatrixcity;
    \addplot[color=blue, mark=*, mark size=0.6pt]
    table[x={distance}, y={vg_count}] from \DataMatrixcity;

    
    \nextgroupplot[
        title = FPS / Distance,
        xmin = -1.0,
        xmax = 20.0,
        xtick distance = 4,
        ymin = 0,
        ymax = 150,
        ymajorgrids=true,
        grid style=loosely dashed,
    ]
    \addplot[color=orange, mark=*, mark size=0.6pt]
    table[x={distance}, y={gs_fps}] from \DataMatrixcity;
    \addplot[color=blue, mark=*, mark size=0.6pt]
    table[x={distance}, y={vg_fps}] from \DataMatrixcity;
    
    
    \nextgroupplot[
        title = FLIP / Distance,
        xmin = -1.0,
        xmax = 20.0,
        xtick distance = 4,
        ymin = 0,
        ymax = 0.0500,
        ytick distance = 0.0125,
        scaled ticks=false,
        y tick label style={/pgf/number format/fixed, /pgf/number format/precision=4, /pgf/number format/fixed zerofill},
        ymajorgrids=true,
        grid style=loosely dashed,
    ]
    \addplot[color=orange, mark=*, mark size=0.6pt]
    table[x={distance}, y={gs_flip}] from \DataMatrixcity;
    \addplot[color=blue, mark=*, mark size=0.6pt]
    table[x={distance}, y={vg_flip}] from \DataMatrixcity;


    \nextgroupplot[
        ylabel = \textsc{Downtown},
        xmin = -1.0,
        xmax = 20.0,
        ymin = 0,
        ymax = 20000000,
        ymajorgrids = true,
        grid style=loosely dashed,
    ]
    \addplot[color=orange, mark=*, mark size=0.6pt]
    table[x={distance}, y={gs_count}] from \DataDowntown;
    \addplot[color=blue, mark=*, mark size=0.6pt]
    table[x={distance}, y={vg_count}] from \DataDowntown;

    
    \nextgroupplot[
        xmin = -1.0,
        xmax = 20.0,
        xtick distance = 4,
        ymin = 0,
        ymax = 150,
        ymajorgrids=true,
        grid style=loosely dashed,
        legend style={
            draw=none,
            anchor=north,   
            at={(0.5,-0.3)},
            legend columns=-1,
            /tikz/every even column/.append style={column sep=1.0em},
        },
    ]
    \addplot[color=orange, mark=*, mark size=0.6pt]
    table[x={distance}, y={gs_fps}] from \DataDowntown;
    \addlegendentry{3DGS}
    \addplot[color=blue, mark=*, mark size=0.6pt]
    table[x={distance}, y={vg_fps}] from \DataDowntown;
    \addlegendentry{Ours}

    
    \nextgroupplot[
        xmin = -1.0,
        xmax = 20.0,
        xtick distance = 4,
        ymin = 0,
        ymax = 0.0500,
        ytick distance = 0.0125,
        scaled ticks=false,
        y tick label style={/pgf/number format/fixed, /pgf/number format/precision=4, /pgf/number format/fixed zerofill},
        ymajorgrids=true,
        grid style=loosely dashed,
    ]
    \addplot[color=orange, mark=*, mark size=0.6pt]
    table[x={distance}, y={gs_flip}] from \DataDowntown;
    \addplot[color=blue, mark=*, mark size=0.6pt]
    table[x={distance}, y={vg_flip}] from \DataDowntown;

    \end{groupplot}
\end{tikzpicture}

\caption{\label{fig:experiment-single-large-scenes-graph}
    \textbf{Graphs on individual large scenes.}
    We report metrics comparing renders of 3DGS and our system on two large scenes at twenty relative distances.
}
\Description{experiment-single-large-scenes-graph}
\end{figure*}

We have already presented results for four composed scenes. Here, we provide additional results for two large scenes, \textsc{Matrixcity} and \textsc{Downtown}, each containing a single reconstructed asset with approximately 10 million 3D Gaussians. The online selection stage uses $\tau = 32768$.

As shown in Fig.~\ref{fig:experiment-single-large-visual} and Fig.~\ref{fig:experiment-single-large-scenes-graph}, our method reduces the percentage of selected 3D Gaussians with increasing distances, accelerating the rendering process. This demonstrates that our approach is effective not only for multiple small assets but also for large assets during both offline build and online selection stages.

However, while these scenes contain around 10 million 3D Gaussians, they are less intricate than object-level assets, leading to blurring effects when viewed up close. For improved quality, it would be beneficial to reconstruct each building separately in high detail and then compose them, enabling better flexibility and LOD control during scene navigation.


\subsection{Results Using Original Views of NVS Datasets\label{sec:appendices-experiments-single-assets}}

We present the average metrics rendered by our proposed system, following the configurations in Sec.~\ref{sec:experiments-implementation-details} and Sec.~\ref{sec:experiments-metrics}, on the original views of NVS datasets used to reconstruct all 3DGS assets.
Footprint tolerances $\tau$s are adjusted to match the original resolution of each image.

As shown in Table~\ref{tab:appendices-datasets-assets-nvs-metrics}, the PSNR values of our system exhibit minimal variation compared to 3DGS, indicating no reduction in quality for the original views. Moreover, the build duration for each 3DGS asset is comparable to the reconstruction time required for 3DGS.


\end{document}